\documentclass{aa}

\usepackage{amsmath,amsfonts,amssymb} %for math
\usepackage{bm} %for vectors
\usepackage[table,  dvipsnames]{xcolor} %for author comment colors and that table
\usepackage{hyperref} %to edit citation properties
\usepackage{nicefrac} %for some nice inline fractions
\usepackage{placeins}
\usepackage{microtype}
\usepackage{siunitx}

\usepackage[varg]{txfonts} %aa requirement
\usepackage[T1]{fontenc}

\usepackage{soul}

\hypersetup{
	pdftitle = {Planet formation at the inner disk edge: the interplay between accretion outbursts and dust growth},
	pdfauthor = {A. Ziampras, T. Birnstiel, N. Kaufmann, M. Cecil, T. Pfeil},
	pdfsubject = {},
	colorlinks = {true},
	linkcolor = [rgb]{0,0.35,0.7},
	citecolor = [rgb]{0,0.35,0.7},
	filecolor = [rgb]{0.61,0,0},
	urlcolor = [rgb]{0.61,0,0},
}

\usepackage{graphicx}  %for images
\graphicspath{{./}} % submission version

\newcommand{\tensorGR}[1]{\overline{\bm{{#1}}}}
\newcommand{\DP}[2]{\frac{\partial{#1}}{\partial{#2}}}
\newcommand{\D}[2]{\frac{\text{d}{#1}}{\text{d}{#2}}}

\newcommand{\G}{\text{G}}
\newcommand{\Mstar}{M_\star}
\newcommand{\Lstar}{L_\star}

\newcommand{\Msun}{\mathrm{M}_\odot}
\newcommand{\Lsun}{\mathrm{L}_\odot}

\newcommand{\Mearth}{\mathrm{M}_\oplus}
\newcommand{\Rgas}{\mathcal{R}}
\newcommand{\cs}{c_\mathrm{s}}

\newcommand{\OmegaK}{\Omega_\mathrm{K}}

\newcommand{\tauR}{\tau_\mathrm{R}}
\newcommand{\tauP}{\tau_\mathrm{P}}

\newcommand{\taueff}{\tau_\mathrm{eff}}
\newcommand{\kappaR}{\kappa_\mathrm{R}}
\newcommand{\kappaP}{\kappa_\mathrm{P}}

\newcommand{\rhomid}{\rho_\mathrm{mid}}
\newcommand{\sigmaSB}{\sigma_\mathrm{SB}}
\newcommand{\vel}{\bm{u}}

\newcommand{\Rrim}{R_\mathrm{rim}}

\newcommand{\Qvisc}{Q_\mathrm{visc}}
\newcommand{\Qcool}{Q_\mathrm{cool}}
\newcommand{\Qsurf}{Q_\mathrm{surf}}
\newcommand{\Qirr}{Q_\mathrm{irr}}
\newcommand{\Qrad}{Q_\mathrm{rad}}

\newcommand{\Sigmag}{\Sigma_\mathrm{g}}
\newcommand{\Sigmad}{\Sigma_\mathrm{d}}
\newcommand{\Sigmadi}{\Sigma_{\mathrm{d},i}}
\newcommand{\velg}{\vel_\mathrm{g}}
\newcommand{\veld}{\vel_\mathrm{d}}
\newcommand{\veldi}{\vel_{\mathrm{d},i}}
\newcommand{\epsi}{\epsilon_i}
\newcommand{\St}{\mathrm{St}}

\newcommand{\amin}{a_\mathrm{min}}
\newcommand{\aint}{a_\mathrm{int}}
\newcommand{\amax}{a_\mathrm{max}}
\newcommand{\Sigmasmall}{\Sigma_\mathrm{small}}
\newcommand{\Sigmalarge}{\Sigma_\mathrm{large}}
\newcommand{\qdust}{q_\mathrm{dust}}
\newcommand{\TMRI}{T_\text{MRI}}
\newcommand{\alphaMRI}{\alpha_\text{MRI}}
\newcommand{\alphaDZ}{\alpha_\text{DZ}}
\newcommand{\Tsubl}{T_\text{subl}}
\newcommand{\Mdotacc}{\dot{M}_\text{acc}}
\newcommand{\Lacc}{L_\text{acc}}

\newcommand{\pluto}{\texttt{PLUTO}}

\newcommand{\optool}{\texttt{optool}}

\newcommand{\tripod}{\texttt{TriPoD}}
\newcommand{\kabs}{\kappa_{\text{abs}}}
\newcommand{\ksca}{\kappa_{\text{sca}}}

\defcitealias{pfeil-etal-2024}{P+24}
\newcommand{\pfeilt}{\citetalias{pfeil-etal-2024}}
\newcommand{\pfeilp}{\citepalias{pfeil-etal-2024}}
\defcitealias{cecil-flock-2024}{CF24}
\newcommand{\cecilt}{\citetalias{cecil-flock-2024}}

\begin{document}
\title{Planet formation at the inner edge of the dead zone}
\subtitle{I. the interplay between accretion outbursts and dust growth}
\titlerunning{Accretion outbursts and dust growth}

\author{
	Alexandros~Ziampras\thanks{E-mail: \texttt{a.ziampras@lmu.de}}\inst{\ref{LMU},\ref{MPIA}}
	\and Tilman~Birnstiel\inst{\ref{LMU},\ref{ORIGINS}}
	\and Nicolas~Kaufmann\inst{\ref{LMU}}
	\and Michael~Cecil\inst{\ref{MPIA}}
	\and Thomas~Pfeil\inst{\ref{CCA}}
}

\institute{
	Ludwig-Maximilians-Universit{\"a}t M{\"u}nchen, Universit{\"a}ts-Sternwarte, Scheinerstr.~1, 81679 M{\"u}nchen, Germany\label{LMU}
    \and Max Planck Institute for Astronomy, K{\"o}nigstuhl 17, 69117 Heidelberg, Germany\label{MPIA}
	\and Exzellenzcluster ORIGINS, Boltzmannstr.~2, 85748 Garching, Germany\label{ORIGINS}
	\and Center for Computational Astrophysics, Flatiron Institute\thanks{The Flatiron Institute is a division of the Simons Foundation.}, 162 5th Ave, New York, NY 10010, USA\label{CCA}
}

\date{\today}

\abstract
	{~The inner edge of the dead zone in protoplanetary disks has been shown to periodically go unstable, leading to accretion outbursts and annular substructure within the dead zone. While dust opacities play a key role in this process, the thermal and dynamical effects of dust drift and growth have not been fully explored.
	}
	{~We investigate the evolution of accretion outbursts in the inner disk and their impact on the formation of dust-rich substructure with a fully dynamic dust model.
	In doing so, we aim to highlight the importance and limitations of dust growth in forming planets in this region.
	}
	{~We carry out radiation hydrodynamics simulations of a protoplanetary disk including prescriptions for the structure of the inner edge of the dead zone, viscous and irradiation heating, radiative cooling, dust--gas dynamics, and dust evolution.
	}
	{~We find that accretion outbursts at the inner disk edge can lead to the formation of multiple dust rings that extend deep inside the dead zone ($\sim\!1$\,au) and diffuse on viscous timescales ($\sim\!10^4$\,yr for $\alpha_\mathrm{DZ}=10^{-4}$). The rings contain dust masses of up to $\sim\!1.6\,\Mearth$, possibly kickstarting planet formation. Dynamic modeling of dust fragmentation enhances the total opacity during the burst, yielding more intense outbursts that penetrate deeper into the dead zone.
	}
	{~Our results highlight the thermal and dynamical importance of treating dust dynamics self-consistently in models of accretion outbursts. 
	Additional modeling is needed to characterize the inevitable nonaxisymmetric structures arising from accretion outbursts and their observational prospects.}

\keywords{accretion discs --- hydrodynamics --- radiation: dynamics --- methods: numerical}

\bibpunct{(}{)}{;}{a}{}{,}

\maketitle

\section{Introduction}
\label{sec:introduction}

% The inner rim has been studied for ages
%The inner regions of protoplanetary disks have been the subject of decades of study, both observationally and theoretically.
Observations of young stellar objects have revealed that many systems exhibit strong variability in their luminosity, often attributed to episodic accretion events \citep{herbig-1977,audard-etal-2014}. Of particular interest are the so-called FU Orionis-type outbursts \citep[e.g.,][]{hartmann-kenyon-1996}, characterized by sudden increases in brightness by several magnitudes that can last for decades to centuries \citep[for a review, see][]{fischer-etal-2023}.

% stability of the inner edge
Theoretical models have proposed various mechanisms to explain these outbursts, with a prominent candidate being a thermal instability (TI) operating at the interface between the inner edge of the dead zone and the ionized active interior \citep{dullemond-monnier-2010}. The latter region is thought to be sufficiently hot and ionized to sustain turbulence via the magnetorotational instability \citep[MRI,][]{balbus-hawley-1991,hawley-etal-1995}, with the dead zone being colder and significantly less turbulent due to insufficient ionization quenching the MRI \citep{gammie-1996,bai-stone-2013b}, albeit with some weak level of hydrodynamic turbulence \citep[for a review, see][]{lesur-etal-2023b}.

\begin{figure}
	\centering
	\includegraphics[width=\columnwidth, trim={0 0 6cm 0}, clip]{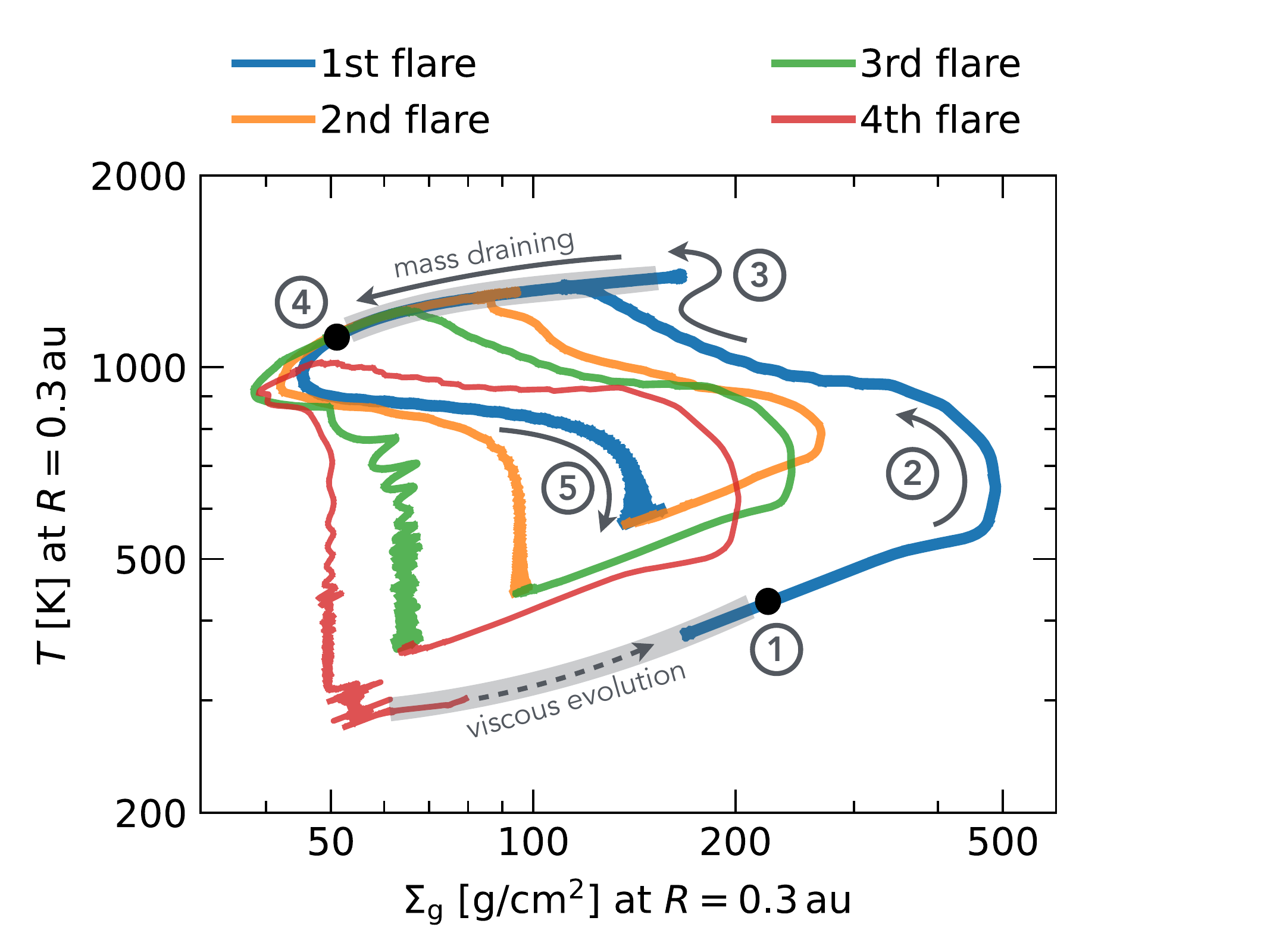}
	\caption{``S-curve'' of the gas surface density versus temperature at $R=0.3$\,au for our fiducial model, discussed in detail in Sect.~\ref{sub:results-fiducial-model}---1: runaway heating is triggered at the DZIE; 2: burst front passes through; 3: dust sublimation caps temperature during burst; 4: dust recondensation; 5: dust cooling. Gray bands highlight the stable branches corresponding to viscous evolution in the quiescent (bottom) and burst (top) phases. Different colors denote reflares during the same burst cycle.}
	\label{fig:fiducial-s-curve}
\end{figure}

This contrast in turbulent angular momentum transport at the dead zone inner edge (DZIE) leads to steady mass accumulation there \citep{flock-etal-2017a}, until the TI is triggered when turbulent dissipation (i.e., viscous heating) exceeds radiative cooling even for the trace amounts of turbulence in the dead zone. In turn, this results in a runaway heating episode, which subsequently ionizes the gas, triggering the MRI and inducing a burst of viscous accretion onto the central star \citep{armitage-etal-2001,zhu-etal-2009,cecil-flock-2024}. An example of a burst cycle via this process is shown in Fig.~\ref{fig:fiducial-s-curve}, using an ``S-curve'' representation of the thermal equilibrium states of the disk \citep[e.g.,][]{faulkner-etal-1983}.

The evolution of an accretion burst via this variant of the TI has been the topic of several numerical studies. Early works by \citet{armitage-etal-2001} and \citet{zhu-etal-2009} used one-dimensional (1D) viscous disk models with simplified thermal physics to demonstrate the viability of this mechanism, while \citet{chambers-2024} used more sophisticated 1D cooling prescriptions recovering similar results. More recent studies have employed two-dimensional (2D) radiation hydrodynamics simulations to capture the complex interplay between heating, cooling, and angular momentum transport at the DZIE \citep{wunsch-etal-2006,cecil-flock-2024}, revealing the presence of reflares within a burst cycle and the formation of multiple rings of high gas density within the dead zone ($\sim\!1$\,au) as a byproduct of the outburst. Such features were also found by \citet{chambers-2024}.

This substructure is particularly interesting from the perspective of planet formation, with the pressure maxima associated with the rings acting as traps for inward-drifting dust particles \citep{weidenschilling-1977,takeuchi-lin-2002}. The accumulation of dust in these regions could facilitate the formation of planetesimals via the streaming instability \citep{youdin-goodman-2005,johansen-youdin-2007,baronett-etal-2024,aly-paardekooper-2025} and their subsequent growth via either pebble accretion \citep[e.g.,][]{lambrechts-johansen-2012,lambrechts-johansen-2014} or planetesimal accretion \citep[e.g.,][]{pollack-etal-1996,weidenschilling-2000}. If such a formation pathway is viable, it could explain the presence of thousands of super-Earths and mini-Neptunes found within $\sim\!1$\,au of their host stars\footnote{\url{https://exoplanetarchive.ipac.caltech.edu/}} \citep{fressin-etal-2013,petigura-etal-2013,petigura-etal-2018} via essentially in situ formation.

With the dust playing a central role in the thermodynamics of accretion outbursts by being the primary carriers of opacity in the dead zone, it is crucial to treat dust dynamics self-consistently when modeling this process. However, previous works on the subject have assumed a fixed dust size distribution of grains perfectly coupled to the gas, with the dust-to-gas ratio being an input parameter rather than a dynamically evolving quantity through radial drift, accumulation at pressure bumps, and dust growth. 

In this work, we aim to address this gap by modeling dust as a fully dynamic component that evolves via coagulation and fragmentation, and interacts with the gas both dynamically via drag forces but also thermally via size- and temperature-dependent dust opacities. This will allow us to assess the impact of dust evolution during and after an accretion outburst, as well as the potential for planet formation in the resulting dust rings. %Furthermore, our models will enable us to make predictions for the observability of such rings with future facilities sensitive to the sub-au regions, such as the next-generation Very Large Array \citep[ngVLA,][]{murphy-etal-2018}.

% nomenclature
Below, we will refer to the quiescent state before an outburst as the ``pre-burst'' phase, the outburst itself as a ``burst cycle'' (or simply ``burst''), and the disk state after the burst has subsided as the ``post-burst'' phase. During the long quiescent phase, this post-burst state evolves viscously until the next pre-burst state is reached, where the conditions for a burst are once again satisfied. The burst cycle itself can consist of multiple ``flares'', with the first being the most intense one, and ``reflares'' being subsequent, less intense events within the same burst cycle. We will also use the terms ``dead zone inner edge'' (DZIE) and ``inner rim'' interchangeably. 

In Sect.~\ref{sec:physics-numerics} we describe the theoretical framework used in our work. We then make semi-analytical predictions on the state of the disk before and after a burst in Sect.~\ref{sec:pre-post-constraints}. The results of our radiation hydrodynamical simulations are presented in Sect.~\ref{sec:results-outbursts}, followed by simplified long-term models of dust evolution that examine the prospects for planetesimal formation in Sect.~\ref{sec:results-planet-formation}.
% and synthetic observations with ngVLA in Sect.~\ref{sec:results-ngvla-synthetic-observations}
We discuss our results in Sect.~\ref{sec:discussion}, and conclude in Sect.~\ref{sec:summary}.

\section{Physics and numerics}
\label{sec:physics-numerics}

In this section, we describe the physical and numerical framework used in our study. We begin by listing the relevant hydrodynamical equations and describing the different processes involved. We then describe the numerical setup used in our hydrodynamical simulations in Sect.~\ref{sec:results-outbursts}.

\subsection{Multifluid radiation hydrodynamics}
\label{sub:physics}

We assume a disk of gas with surface density $\Sigmag$, velocity field $\velg$, internal energy density $e$, mean molecular weight $\mu=2.353$, and adiabatic index $\gamma=7/5$, as well as a pressureless dust component with surface density $\Sigmad$ and velocity $\veld$, orbiting a star of mass $\Mstar=1\,\Msun$ and luminosity $\Lstar=1.78\,\Lsun$ \citep{hayashi-1981}. The vertically integrated Navier--Stokes equations then read
\begin{subequations}
	\label{eq:navier-stokes}
	\begin{align}
	\label{eq:navier-stokes-1}
	\D{\Sigmag}{t} &= -\Sigmag\nabla\cdot\velg, \\
	\label{eq:navier-stokes-2}
	\D{\velg}{t} &= -\frac{1}{\Sigmag} \nabla P - \nabla \Phi_\star + \nabla\cdot\tensorGR{\sigma} - \sum\limits_{i}\epsi \frac{\veldi - \velg}{\St_i}\OmegaK, \\
	\label{eq:navier-stokes-3}
	\D{e}{t} &= -\gamma e\nabla\cdot\velg + \Qvisc + \Qirr + \Qcool + \Qrad, \\
	\label{eq:navier-stokes-4}
	\D{\Sigmadi}{t} &= -\Sigmadi\nabla\cdot\veldi - \nabla\cdot\bm{F}_{\mathrm{diff},i},\quad \bm{F}_{\mathrm{diff},i} = -\Sigmag\frac{\nu}{1+\St_i^2}\nabla\epsi\\
	\label{eq:navier-stokes-5}
	\D{\veldi}{t} &= -\nabla \Phi_\star - \frac{\veldi - \velg}{\St_i}\OmegaK.
	\end{align}
\end{subequations}
In the above equations, $\text{d}/\text{d}t$ refers to the material derivative, $P=(\gamma-1)e$ is the gas pressure given by the perfect gas law, $\Phi_\star = -\G\Mstar/R$ is the gravitational potential of the star at distance $R$, $\tensorGR{\sigma}$ is the viscous stress tensor, $\OmegaK=\sqrt{\G\Mstar/R^3}$ is the Keplerian angular frequency, and $\nu$ is the kinematic viscosity of the gas. The gravitational constant is denoted by $\G$. Through the above we can define the isothermal sound speed $\cs=\sqrt{P/\Sigmag}$, the temperature $T=\mu\cs^2/\Rgas$, the pressure scale height $H=\cs/\OmegaK$, and the aspect ratio $h=H/R$, with $\Rgas$ being the ideal gas constant.

Regarding the dust component for a species with index $i$, we define the dust-to-gas ratio $\epsi=\Sigmadi/\Sigmag$ and the Stokes number (i.e., dimensionless stopping time) of a species with radius $a_i$ and material density $\tilde{\rho}=2.08\,\text{g}/\text{cm}^3$ in the Epstein regime as
\begin{equation}
\St_i = \frac{\pi}{2}\frac{a_i\tilde{\rho}}{\Sigmag}.
\end{equation}
The term $\bm{F}_{\mathrm{diff},i}$ then represents dust diffusion due to turbulence \citep{morfill-voelk-1984,youdin-lithwick-2007}. 

The terms $\Qvisc$, $\Qirr$, $\Qsurf$, and $\Qrad$ in Eq.~\eqref{eq:navier-stokes-3} represent viscous heating, stellar irradiation heating, radiative cooling through surface losses, and radiative diffusion through the disk midplane, respectively. Similar to \citet{ziampras-etal-2025c}, we define these terms as follows:
\begin{subequations}
	\label{eq:source-terms}
	\begin{align}
		\label{eq:source-terms-1}
		Q_\mathrm{visc} = \frac{1}{2\nu\Sigmag}\mathrm{Tr}(\tensorGR{\sigma}^2) \approx \frac{9}{4}\nu\Sigmag\OmegaK^2, \qquad \nu=\alpha\sqrt{\gamma}\cs H,
	\end{align}
	\begin{align}
		\label{eq:source-terms-2}
		Q_\mathrm{irr} = 2\frac{\Lstar}{4\pi R^2} (1-\varepsilon)\frac{\theta}{\taueff},
	\end{align}
	\begin{align}
		\label{eq:source-terms-3}
		\Qsurf = -2\frac{\sigmaSB T^4}{\taueff},~~\taueff=\frac{3\tauR}{8}\!+\!\frac{\sqrt{3}}{4}\!+\!\frac{1}{4\tauP},~~\tau_\text{R,P}\!=\!\frac{\kappa_\text{R,P}\Sigmag}{2},
	\end{align}
	\begin{align}
		\label{eq:source-terms-4}
		\Qrad = \sqrt{2\pi}H\nabla\cdot \left(\lambda\frac{4\sigmaSB}{\kappaR\rhomid}\nabla T^4\right), \quad \rhomid = \frac{1}{\sqrt{2\pi}}\frac{\Sigmag}{H}.
	\end{align}
\end{subequations}
In the above, we utilize the $\alpha$-viscosity model of \citet{shakura-sunyaev-1973}, an irradiation model based on \citet{menou-goodman-2004} with a disk albedo $\varepsilon=1/2$ and flaring angle $\theta$, an effective optical depth $\taueff$ following \citet{hubeny-1990}, and in-plane radiative transfer in the one-temperature flux-limited diffusion approximation \citep{levermore-pomraning-1981,rometsch-etal-2024}. The Rosseland and Planck mean opacities are denoted by $\kappaR$ and $\kappaP$, respectively, $\rhomid$ is the midplane gas volume density, $\lambda$ is the flux limiter from \citet{kley-1989}, and $\sigmaSB$ is the Stefan--Boltzmann constant. Our prescriptions specific to the dead zone inner edge for $\alpha$, $\theta$, $\kappaR$, and $\kappaP$ are described in Sects.~\ref{sub:dust-model}~\&~\ref{sub:physics-rim}.

We note here that in-plane radiative diffusion via $\Qrad$ is only included in a subset of our simulations, due to its steep computational cost during the post-burst phase. Its effects and relevance are explored in detail in Sect.~\ref{sub:results-tripod-FLD}.

\subsection{Dust evolution and opacity model}
\label{sub:dust-model}

In addition to the set of equations in Eq.~\eqref{eq:navier-stokes}, we also solve for the evolution of the dust size distribution $n(a)$ using the \tripod{} method \citep[][hereafter \pfeilt]{pfeil-etal-2024}. This approach simplifies the full coagulation equations \citep{smoluchowski-1916} by instead tracking the evolution of representative ``small'' and ``large'' grain populations with surface densities $\Sigmasmall$ and $\Sigmalarge$ via coagulation and/or fragmentation, along with the maximum grain size $\amax$ \citepalias[following][we fix $\amin=0.1\,\mu$m]{pfeil-etal-2024}. This method can be seen as a more accurate extension of \texttt{two-pop-py} \citep{birnstiel-etal-2017}, with a wider range of applicability. For the full documentation, implementation, and testing of the \tripod{} method we refer the reader to \pfeilt.

\begin{figure}
	\centering
	\includegraphics[width=\columnwidth]{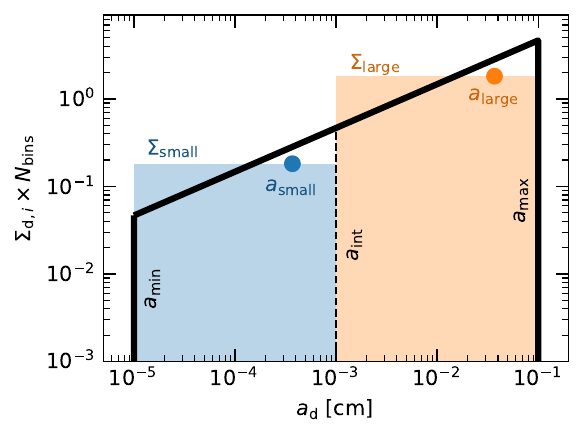}
	\caption{Sample dust size distribution, represented with a truncated power-law with $\qdust=-3.5$ and reconstructed with the \tripod{} method of \citet{pfeil-etal-2024}.}
	\label{fig:tripod}
\end{figure}

In the \tripod{} approach, $\Sigmalarge$ and $\Sigmasmall$ define a power-law dust size distribution, truncated at a maximum grain size $\amax$. Given the two populations are divided at an intermediate grain size $\aint = \sqrt{\amin\,\amax}$, they define the power-law exponent of the distribution $\qdust$ via
\begin{equation}
	\label{eq:qdust}
	\qdust = \frac{\log(\Sigmalarge/\Sigmasmall)}{\log(\amax/\aint)} - 4, 
\end{equation}
such that $\Sigmadi(a)\propto a^{\qdust+4}$. 
The cutoff size $\amax$ is treated as an additional ``fluid'' that is advected with the large population's velocity field, that is, tracing the advection of the largest grains. Figure~\ref{fig:tripod} illustrates the reconstructed dust size distribution using $\Sigmasmall$, $\Sigmalarge$, and $\amax$.

Since dust is modeled as a set of fluids with their own velocities in our \pluto{} implementation, we further account for momentum conservation during coagulation and fragmentation by exchanging momenta between $\Sigmasmall$ and $\Sigmalarge$ during the \tripod{} step. This is not the case in the original \texttt{TriPoD} implementation of \citetalias{pfeil-etal-2024}. Simple tests showed that this does not affect the results in any meaningful way, so a recalibration of the method was not necessary, but we include this detail for completeness.

Regarding our opacity model, we can write the total Rosseland or Planck opacity $\kappa$ in Eqs.~\eqref{eq:source-terms-3}~\&~\eqref{eq:source-terms-4} as the sum of dust and gas opacities $\kappa_\text{d}$ and $\kappa_\text{g}$, expressed in $\text{cm}^2$ per gram of gas:
\begin{equation}
	\label{eq:opacity}
	\kappa = f_\text{subl}\,\epsilon\,\kappa_\text{d} + \kappa_\text{g},\qquad\epsilon=\frac{\Sigmasmall+\Sigmalarge}{\Sigmag},
\end{equation}
with $f_\text{subl}$ being a prefactor that accounts for dust sublimation at high temperatures (see Eq.~\eqref{eq:fsubl} in Sect.~\ref{sub:physics-rim}). For our dust opacities we use the $\amax$-, $\qdust$-, and $T$-dependent tables computed with the \texttt{growpacity}\footnote{\url{https://github.com/alexziab/growpacity}} module \citep{growpacity} and fix $\kappa_\mathrm{R,gas}=\kappa_\mathrm{P,gas} = 10^{-3}\,\text{cm}^2/\text{g}_\text{gas}$ following \citet{cecil-flock-2024} \citepalias[hereafter][]{cecil-flock-2024}. A summary of the \texttt{growpacity} method is provided in Appendix~\ref{apdx:growpacity}.

\subsection{Model adaptation to the dead zone inner edge}
\label{sub:physics-rim}

Interior to the dead zone inner edge, the MRI is thought to operate at full strength, driving intense levels of turbulence \citep{flock-etal-2017a,iwasaki-etal-2024}. Following \cecilt, we model the MRI activity as a temperature-dependent $\alpha$, under the assumption that the gas is sufficiently ionized above a temperature $\TMRI$ such that the MRI can operate:
\begin{equation}
	\label{eq:alpha-mri}
	\alpha(T) = \alphaDZ + \frac{1}{2}\left(\alphaMRI-\alphaDZ\right)\left[1 + \tanh\left(\frac{T-\TMRI}{\Delta T_\mathrm{MRI}}\right)\right],
\end{equation}
with $\alphaMRI=0.1$, $\TMRI=900$\,K, and $\Delta T_\mathrm{MRI}=25$\,K. Similarly, we model the sublimation of dust grains beyond a critical temperature $\Tsubl = 2000\,\text{K}\,\rho_{\text{mid,g/cm}^3}^{0.0195}$ \citep{isella-natta-2005} with a sublimation fraction given similar to \cecilt{}:
\begin{equation}
	\label{eq:fsubl}
	f_\text{subl} = 10^{-10} + \frac{1}{2}\left(1-10^{-10}\right)\left[1 - \tanh\left(\frac{T-\Tsubl}{\Delta \Tsubl}\right)\right],
\end{equation}
with $\Delta \Tsubl = 100$\,K.

We note that, in our model, the sublimation fraction above does not directly affect the dust densities $\Sigmadi$, and sublimated dust vapor is allowed to instantly recondense as temperatures drop below $\Tsubl$. This simplification does not hurt the accuracy of our model, however, as the temperatures needed to sublimate the dust practically always satisfy $\Tsubl > \TMRI$, translating to extreme values of $\alpha$ that lead to rapid fragmentation of grains. As a consequence to that, the typical grain sizes during this state are very low ($\sim\!\mu$m), and the ``dust vapor" has a Stokes number so small that it is practically coupled to the background gas flow, effectively evolving as a ``gas'' fluid. As the recondensed dust grows back to larger grains via coagulation, the dust population smoothly decouples from the gas once again.

Finally, to model the direct irradiation of the dust-free, hot inner disk interior to the dead zone edge, we modify the grazing angle in Eq.~\eqref{eq:source-terms-2} from its usual profile of $\theta_0\approx2\chi h/7$ with $\chi=4$ \citep{chiang-goldreich-1997} to reach a value of 1 (i.e., direct illumination) in the dust-free region below $\Rrim=0.1$\,au. This value for the dead zone inner edge was chosen matching the results of \cecilt, and the transition is modeled similarly to above as
\begin{equation}
	\label{eq:theta-rim}
	\theta(R) = \theta_0 + \frac{1}{2}\left(1 - \theta_0\right)\left[1 - \tanh\left(\frac{R - \Rrim}{0.01\,\text{au}}\right)\right].
\end{equation}
This results in a significantly hotter region interior to $\Rrim$ that always satisfies that $\alpha=\alphaMRI$ through Eq.~\eqref{eq:alpha-mri}, and ensures that the disk has a cavity within $\Rrim$ (see, e.g., Fig.~\ref{fig:fiducial-Rt}). We note that more complex and realistic models of the temperature profile near the inner rim exist \citep[e.g.,][]{ueda-etal-2017}, taking into account the full vertical structure of the disk, but this approach is sufficient for the purposes of this work as we are focused at the disk midplane.

\subsection{Numerics}
\label{sub:numerics}

For our hydrodynamical simulations in Sect.~\ref{sec:results-outbursts} we use the finite-volume \pluto{} \texttt{v.4.4} code \citep{mignone-etal-2007} with the HLLC Riemann solver \citep{toro-etal-1994}, the flux limiter by \citet{vanleer-1974}, and second-order spatial and temporal accuracy. The modules for FLD, dust--gas interaction, and dust growth via \tripod{} are documented in \citet{ziampras-etal-2020a}, \citet{ziampras-etal-2025a}, and \pfeilt, respectively.

We use an axisymmetric cylindrical polar grid that extends radially between 0.05--10\,au, with 1024 cells between 0.05--2\,au and 128 beyond 2\,au, for a total of 1152 cells. Both grid segments stretch outwards with logarithmic spacing. We use a strict outflow boundary condition at the inner boundary for all quantities, and apply a damping zone in the region beyond 8.5\,au towards the initial conditions using the formula in \citet{devalborro-etal-2006} and a damping timescale of 0.1 local orbits.

We typically initialize the disk with a surface density profile that is either nearly or certainly unstable to an accretion outburst, and a power-law temperature profile that corresponds to a passively irradiated disk with $T_0\approx110\,R_\text{au}^{-3/7}\,$K that nevertheless quickly converges to the equilibrium set by $\Qvisc+\Qirr=\Qcool$ in regions where viscous heating or direct irradiation matter. To facilitate a comparison among our models with different $\alphaDZ$, we impose that the surface density in the outer disk is always $\Sigma_\text{disk,0} = 200\,R_\text{au}^{-15/14} \text{g/cm}^2$. The gas velocity is then initialized near viscous and hydrodynamical equilibrium with $u_R=-1.5\nu/R$ and $u_\phi=R\OmegaK\sqrt{1-3h^2/2}$.

For the dust, we initialize both dust densities to $0.005\,\Sigma_0$ for a total dust-to-gas ratio $\epsilon=0.01$, the maximum grain size at a uniform $\amax=1\,$mm, and the dust velocities to $u_{R,i}=0$ and $u_{\phi,i}=R\OmegaK$. As with the gas, these choices are largely inconsequential as dust drift and evolution will quickly adjust the dust size distribution to its equilibrium profile before a burst cycle, but we list them for completion.

Similar to \cecilt, we treat our model with $\alphaDZ=10^{-3}$ as our fiducial setup and run additional models with $\alphaDZ\in[3\times10^{-4}, 10^{-4}]$ to investigate the role of diffusion and dust evolution during and shortly after the burst phase. To highlight the impact of dynamic dust evolution with \tripod{} we additionally run a model where we disable dust growth/fragmentation but instead prescribe a radial profile of $\amax$ and $\qdust$ informed by our fiducial model. Finally, we note that due to very high computational costs during the quiescent phase we opted to not include in-plane radiative diffusion ($\Qrad$, Eq.~\eqref{eq:source-terms-4}) in most of our models, but we nevertheless carry out a run where $\Qrad$ is included during the burst phase to showcase its effect and relevance during and after the burst. %Table~\ref{table:models} summarizes the models that we executed with \pluto{} in this work.

For our results with $\alphaDZ=10^{-3}$ in Sect.~\ref{sub:results-fiducial-model} we simply initialize the gas surface density with $\Sigma_0=\Sigma_\text{disk,0}$. This will result in an artificially more massive burst, but viscous timescales are short enough that we can discard this first burst cycle and evolve until the next burst self-consistently triggers, which we then analyze. For our models with $\alphaDZ\leq3\times10^{-4}$ these viscous timescales are prohibitively long, and we therefore initialize the gas surface density with the pre-burst state predicted using the method described in Sect.~\ref{sec:pre-post-constraints} and tested in Appendix~\ref{apdx:pre-post-burst}.

\section{Constraints on the pre- and post-burst states}
\label{sec:pre-post-constraints}

Here, we will use the physical framework described in Sect.~\ref{sub:physics} to derive approximate solutions to the pre- and post-burst states near the inner edge of the dead zone. We will also provide illustrative examples of these solutions.

\subsection{Pre-burst state}
\label{sub:pre-burst}

As shown by the aforementioned studies \citep[e.g.,][]{armitage-etal-2001,zhu-etal-2009, cecil-flock-2024}, an accretion burst is triggered once sufficient material has accumulated at the dead zone inner edge such that the viscous heating term $\Qvisc\propto\Sigmag$ can overcome the radiative cooling term $\Qcool\propto\Sigmag^{-1}$ (for $\tau\gg1$, which is easily satisfied in the optically thick inner disk). This triggers a runaway heating phase where $\Qvisc \propto \alpha(T)$ further increases, until the temperature reaches $T\approx\Tsubl$ and is regulated by a ``thermostat effect'': a drop in temperature would allow dust to recondense and thus the opacity to increase, promoting heating, whereas a further increase to temperature is not feasible due to the lack of a dust opacity and a very low gas opacity \citep[see, however][for the role of realistic gas opacities in the disk atmosphere]{cecil-etal-2026}.
During this ``hot'' state, the gas is evolving at viscous rates with $\alpha\approx\alphaMRI$, with a substantial fraction of the gas mass within 1\,au accreted onto the central star by the end of this phase. The remaining gas is rarefied and optically thinner, allowing radiative cooling to once again overpower viscous heating and lowering temperatures back to the quiescent, ``cold'' state.

% \citeme{maybe this can be moved to the introduction and a reference to the S-curve placed here.}

With this picture in mind, we can already place constraints on the shape of the pre- and post-burst states in terms of the gas density and temperature. In particular, during the burst phase, the ``burst front'' traveling outwards cannot exceed the radius beyond which $\Qvisc \leq \Qcool$ even for $\alpha\sim\alphaMRI$. Assuming that the ``hot'' state is characterized by $\Qvisc\gg\Qirr$ and $T\approx\TMRI$, Eq.~\eqref{eq:alpha-mri} yields $\alpha(\TMRI) = \alphaMRI/2$ and we have
\begin{equation}
	\begin{aligned}
	\label{eq:sigma-min}
	\Qvisc(\TMRI) &= \Qcool(\TMRI) \\
	&\Rightarrow \frac{9}{4}\frac{\alphaMRI}{2}\sqrt{\gamma}\cs^2 \Sigma_\text{g,min} \OmegaK = 2\frac{\sigmaSB \TMRI^4}{\frac{3}{8}\frac{1}{2}\kappaR(\TMRI)\Sigma_\text{g,min}}\\
	&\Rightarrow \Sigma_\text{g,min} = \sqrt{\frac{1}{\alphaMRI\OmegaK}\frac{256}{27}\frac{\mu\sigmaSB}{\Rgas\sqrt{\gamma}}\frac{\TMRI^3}{\kappaR(\TMRI)}}.
	\end{aligned}
\end{equation}
In other words, the density must be at least $\Sigma_\text{g,min}$ in order to sustain the outward-traveling heating front, and the burst flare can no longer propagate outwards when this condition is no longer satisfied. As a result, excluding the always-MRI-active region interior to $\Rrim$, the disk surface density profile in the immediate post-burst state can be approximated by
\begin{equation}
	\label{eq:sigma-post}
	\Sigma_\text{g,post} \approx \min\left(\Sigma_\text{g,min}, \Sigma_\text{g,disk}\right),
\end{equation}
where $\Sigma_\text{g,disk}$ is the unaffected density profile of the disk outside of the burst region. We note that, for constant $\alphaMRI$, $\TMRI$, and a temperature-dependent opacity model, Eq.~\eqref{eq:sigma-min} simplifies to
\begin{equation}
    \label{eq:sigma-min-approx}
	\Sigma_\text{g,min} \propto \OmegaK^{-1/2} \propto R^{3/4},
\end{equation}
which provides a very good match to the 3D axisymmetric models of \cecilt{} ($\Sigma_\text{g,min}\propto R^{0.81}$ therein). As we will see in Fig.~\ref{fig:pre-post-examples}, this profile of $\Sigma_\text{g,min}\propto R^{0.81}$ is exactly reproduced in our models due to the dependence of $\kappa$ on $T$ through $f_\text{subl}$ in Eq.~\eqref{eq:fsubl}.

The corresponding temperature profile in the post-burst state can then be computed by solving Eq.~\eqref{eq:navier-stokes-3} in thermal equilibrium, or
\begin{equation}
	\label{eq:q-iter}
	\Qvisc(T') + \Qirr(T') = \Qcool(T')
\end{equation}
with respect to $T'$, using $\Sigma_\text{g,post}$ from Eq.~\eqref{eq:sigma-post}. This can be done numerically via a root-finding algorithm such as Newton--Raphson or bisection.

Equation~\eqref{eq:sigma-post} offers some insight on the radial extent of the disk that can be ``ignited'' during an accretion outburst via this mechanism, with more massive disks being prone to wider regions where the surface density profile can be affected by a burst event. In particular, for our choice of $\alphaMRI$, $\TMRI$, and $\kappaR(\qdust=-3.75, \amax=200\,\mu\text{m},T=\TMRI)$ and for $\Mstar=1\,\Msun$, we find that $\Sigma_\text{g,min} \sim 200\,R_\text{au}^{0.75}\,\text{g/cm}^2$. This would suggest that a Minimum Mass Solar Nebula (MMSN) disk with $\Sigma_\text{g,disk}\approx 1700 \,R_\text{au}^{-3/2}\,\text{g/cm}^2$ \citep{weidenschilling-1977b} would feature an inverted density slope interior to 4 au, whereas the burst-prone region would shrink to within 1\,au for a less massive disk with $\Sigma_\text{g,disk}\approx 200\,R_\text{au}^{-1}\,\text{g/cm}^2$.

\subsection{Post-burst state}
\label{sub:post-burst}

Post-burst, the disk will spend a comparatively longer period in a quiescent state until the conditions for a burst can be fulfilled again. This phase more or less corresponds to the time it takes for the mass in the inner regions affected by the burst to be replenished via viscous accretion from the outer disk, a process happening on timeframes proportional to the viscous timescale with $\alpha=\alphaDZ$. This state is not very exciting from a gas dynamics point of view, as it is largely determined by viscous evolution until enough material has been delivered to the inner disk such that a new burst cycle can begin.

The quiescent phase will last as long as the following condition is satisfied everywhere
\begin{equation}
	\label{eq:stability-criterion}
	\DP{\Qvisc}{T} < \DP{|\Qcool|}{T}.
\end{equation}
In other words, the disk must cool efficiently enough to prevent runaway heating after a small temperature increase, otherwise a new burst cycle will begin, erasing all ``progress'' in terms of mass accumulation in the inner disk. Unfortunately, modeling this would require following the hydro- and thermodynamical evolution of the quiescent phase, which can be quite long for low values of $\alphaDZ$. Nevertheless, assuming that the density evolves entirely viscously, that is \citep{lynden-bell-pringle-1974}
\begin{equation}
	\label{eq:pringle}
	\DP{\Sigmag}{t} = \frac{3}{R}\DP{}{R}\left[\sqrt{R}\DP{}{R}\left(\nu\Sigmag\!\sqrt{R}\right)\right],
\end{equation}
and that the temperature will respond to the density evolution via a balance among viscous heating, irradiation, and thermal cooling on timescales much shorter than the viscous timescale, we can estimate both the duration of the quiescent phase and the next pre-burst state by iterating between a viscous evolution solver that updates $\Sigmag$ and a root-finding algorithm that updates $T$ via Eq.~\eqref{eq:q-iter}, checking if Eq.~\eqref{eq:stability-criterion} is satisfied after every pair of updates\footnote{We note that both Eqs.~\eqref{eq:q-iter}~and~\eqref{eq:pringle} can be written in implicit form for $\Sigmag'$ and $T'$, accelerating this process significantly.}. In Appendix~\ref{apdx:pre-post-burst} we demonstrate that this method produces results that agree very well to a fully radiation-hydrodynamical model from our simulation set in this work.

In all cases, the criterion in Eq.~\eqref{eq:stability-criterion} is first violated at the DZIE ($\sim\!0.1$\,au), which is where the next burst cycle will begin. However, a clear explanation behind the timescale separation between bursts is still lacking from a theoretical point of view and is beyond the scope of this work. Nevertheless, after computing the pre- and post-burst states for three different disk configurations in Fig.~\ref{fig:pre-post-examples}, we find that the quiescent phase lasts for $\sim3$, 20, and 100\,kyr for $\alphaDZ=10^{-3}$, $3\times10^{-4}$, and $10^{-4}$, respectively, suggesting an empirical scaling of $\alpha^{-3/2}$ for the quiescent phase duration. This could be related to the fact that the viscous timescale scales as $t_\text{visc}=R^2/\nu\propto 1/\alpha$, and that low-$\alpha$ models require more mass to accumulate before a burst can occur, to compensate for the weaker viscous heating ($\Qvisc\propto\alpha$). From the latter point, the minimum surface density for a burst to occur then scales as $\Sigma\propto\alpha^{-1/2}$ (see also Eq.~\eqref{eq:sigma-min}, albeit with $\alphaMRI$ being used in that context as it refers to the post-burst state). Combining these scalings yields a quiescent phase duration $\propto \alpha^{-3/2}$, in line with our results.

\subsection{Example pre- and post-burst states}
\label{sub:example-pre-post}

\begin{figure*}
	\centering
	\includegraphics[width=\textwidth]{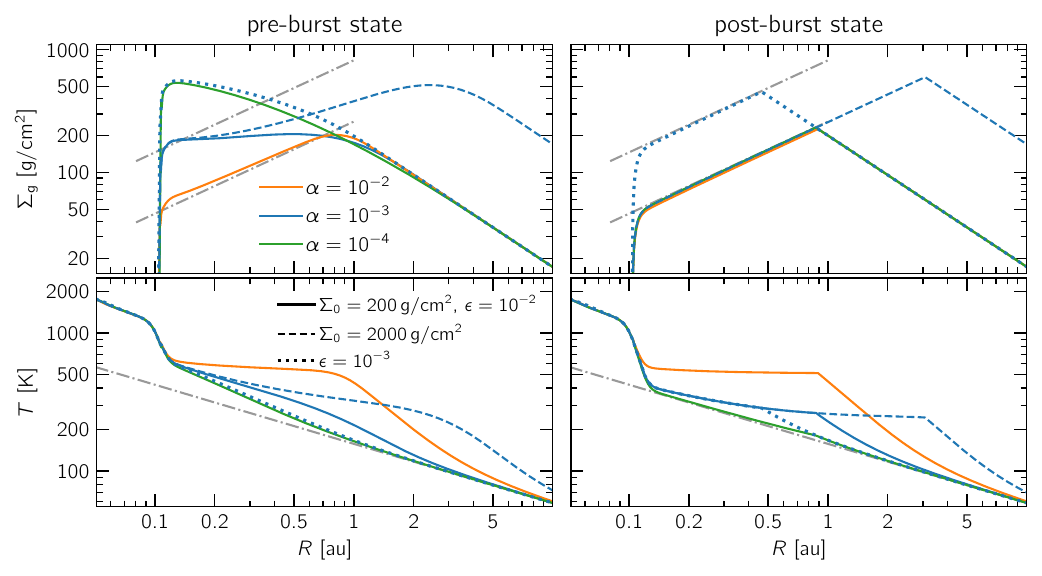}
	\caption{Pre- (left) and post-burst (right) states of the gas surface density (top) and temperature (bottom) for different disk configurations computed using the method described in Sect.~\ref{sec:pre-post-constraints}. Here, we assume constant dust opacities of $\kappa_\text{d}=700\,\text{cm}^2/\text{g}_\text{dust}$ (including $f_\text{subl}$ from Eq.~\eqref{eq:fsubl}). The mass accretion rate through the outer boundary is approximately $10^{-9}\,\Msun/\text{yr}\times\nicefrac{\alphaDZ}{10^{-3}}\times\nicefrac{{\Sigmag}_0}{200\,\text{g}/\text{cm}^2}$. Dashed-dotted lines in the top panels have a slope $\propto R^{0.81}$---in good agreement with Eq.~\eqref{eq:sigma-min-approx}---and help guide the eye, while in the bottom panels they represent the irradiation temperature.}
	\label{fig:pre-post-examples}
\end{figure*}

We can now compute the pre- and post-burst states for different values of $\alphaDZ$ and profiles of $\Sigma_\text{g,disk}$ to showcase a few different disk configurations while accounting for their stability to accretion bursts near the inner edge of their dead zone. Some indicative results are shown in Fig.~\ref{fig:pre-post-examples} for constant dust opacities of $\kappa_\text{d}=700\,\text{cm}^2/\text{g}_\text{dust}$, and different combinations of $\alphaDZ$, $\Sigma_\text{g,disk}$, and $\epsilon$. From the figure we can see how disks with lower $\alphaDZ$ or $\epsilon$ tend to accumulate more mass in the inner regions before bursting, whereas more massive disks have more extended regions affected by the burst. Both trends are in agreement with the physical picture described above.

\section{Results: accretion outbursts}
\label{sec:results-outbursts}

\begin{figure*}
	\centering
	\includegraphics[width=\textwidth]{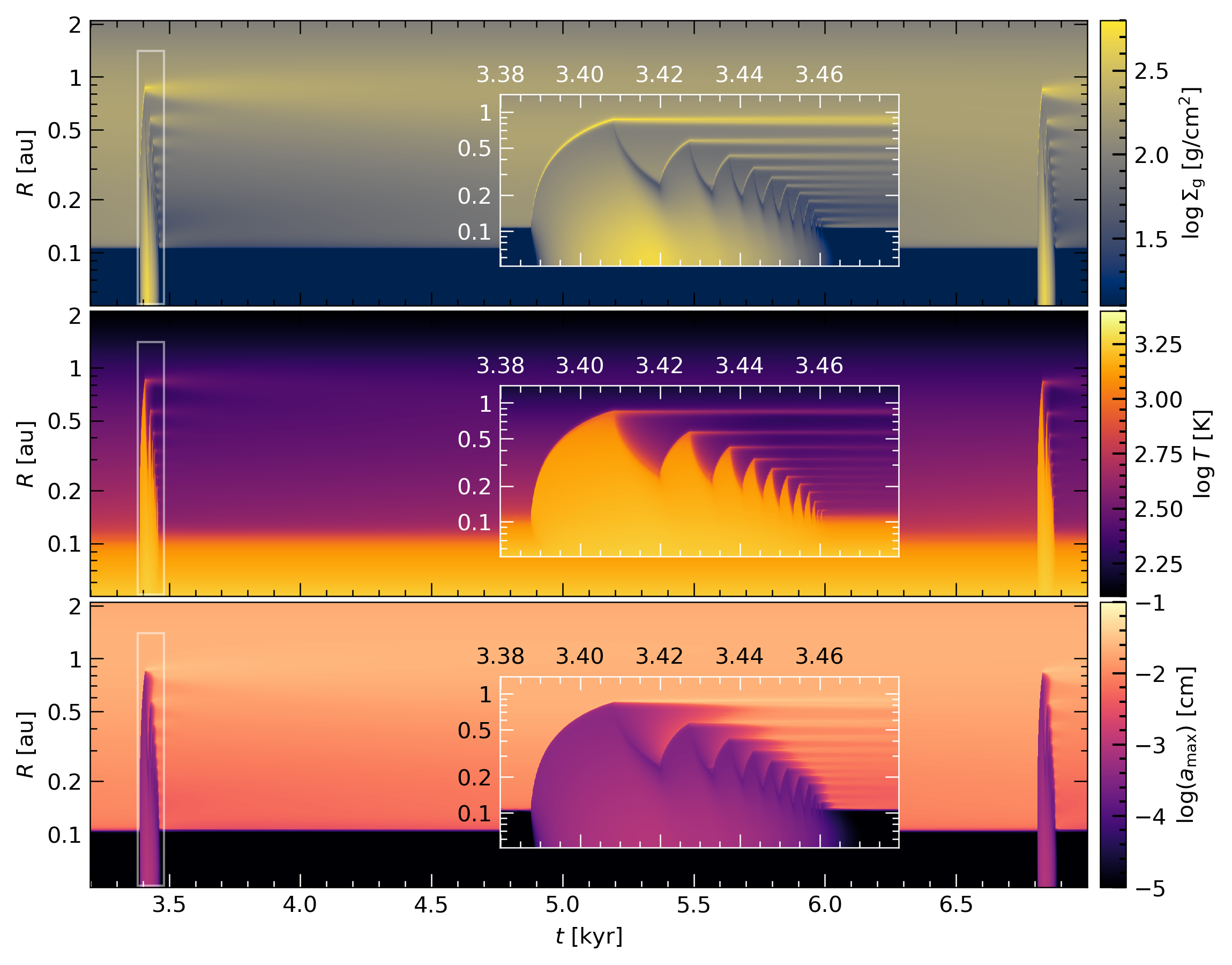}
	\caption{Radius--time heatmaps of the gas surface density (top), temperature (middle), and maximum dust grain size (bottom) as functions of radius for our fiducial model with $\alphaDZ=10^{-3}$. The burst phase is highlighted with insets and marked with white boxes in the main panels, showcasing the emergence of reflares and substructures in the form of rings during the outburst. The rings diffuse away due to viscosity during the long quiescent phase following a burst. The behavior shown here repeats periodically, with the next burst cycle beginning at $\sim\!6.8\,\text{kyr}$ of evolution (see also Fig.~\ref{fig:fiducial-mdot-lacc}).}
	\label{fig:fiducial-Rt}
\end{figure*}

In this section we present the results from our radiation hydrodynamical models of the dead zone inner edge. We begin by describing in detail the results from our fiducial setup in terms of the behavior of the gas and dust during an accretion outburst as well as in the post-burst phase. We then analyze the impact of different thermodynamical prescriptions on the burst characteristics and dust evolution. Finally, we present models with different values of $\alphaDZ$ to showcase the role of viscous diffusion in the burst and post-burst evolution.

\subsection{Fiducial model}
\label{sub:results-fiducial-model}

As viscous evolution is rather quick for $\alphaDZ=10^{-3}$, to the point that we can integrate over multiple burst cycles, we choose this value for our fiducial model. This choice has the additional advantage of more closely mimicking the setup in \cecilt, making a direct comparison easier. For this model, we initialized our disk with power-laws in density with
\begin{equation}
    \label{eq:init-density}
    {\Sigmag}_0 = 200\,R_\text{au}^{-\nicefrac{15}{14}}\,\nicefrac{\text{g}}{\text{cm}^{2}},\quad\!{\Sigmasmall}_0=0.3\epsilon_0{\Sigmag}_0,\quad\!{\Sigmalarge}_0=0.7\epsilon_0{\Sigmag}_0,
\end{equation}
with $\epsilon_0=0.01$ being the initial dust-to-gas ratio. The temperature is initialized with a simple power-law assuming $\Qsurf=\Qirr$ in the dead zone:
\begin{equation}
    \label{eq:init-temperature}
    T_0 = 157\,R_\text{au}^{-3/7}\,\text{K} \Rightarrow h_0 = 0.025\,R_\text{au}^{2/7}.
\end{equation}
Finally, the gas and dust velocities are set to (nearly) Keplerian for this configuration accounting for viscous drift, with
\begin{equation}
    \label{eq:init-velocities}
    {u_R}_\text{g} = {u_R}_\text{d} = -\frac{3}{2}\frac{\nu}{R}, \quad\!{u_\phi}_\text{g} = R\OmegaK\sqrt{1-\nicefrac{3}{2}h^2}, \quad {u_\phi}_\text{d} = R\OmegaK.
\end{equation}
This setup would approximate a constant accretion rate $\dot{M}=-2\pi\Sigmag R{u_R}_\text{g}=3\pi\nu\Sigmag\approx10^{-9}\,\Msun/\text{yr}$ in viscous equilibrium in the dead zone, but of course does not account for the direct illumination of the inner rim (see Eq.~\eqref{eq:theta-rim}), the dependence of $\alpha$ on temperature, or dust growth \citep{birnstiel-etal-2012}. As a result of the former two, it is immediately unstable to the thermal instability near the $\alpha$ transition region at $\Rrim$, and results in an accretion burst, albeit with unrealistic initial conditions. For this reason, similar to \cecilt, we discard the evolution of this first outburst and the viscous evolution that follows it and only analyze data starting from the next burst thereafter, where the dust--gas mixture has had the chance to adapt to the viscously evolving disk, the dust has reached a fragmentation/coagulation equilibrium, and the temperature has adjusted to the thermal equilibrium given by all terms in Eq.~\eqref{eq:navier-stokes-3}.

Figure~\ref{fig:fiducial-Rt} shows an overview of the results of our fiducial model in terms of the gas density $\Sigmag$, temperature $T$, and maximum dust grain size $\amax$ as functions of radius and time. The accretion burst phase is highlighted with insets, revealing the formation of substructures in the form of rings and the presence of reflares similar to those documented in \cecilt. The figure also illustrates the viscous evolution phase of the disk between burst cycles, over which the substructures formed during the burst phase dissolve due to viscous diffusion. Other general features of the inner disk edge are also visible, such as the hot, MRI-active, low-density cavity interior to $\Rrim\approx0.1$\,au and the relative durations of the burst and quiescent phases, lasting $\sim\!85$\,yr and $\sim\!3200$\,yr, respectively. In Appendix~\ref{apdx:fiducial-Rt-extended} we show an extended version of Fig.~\ref{fig:fiducial-Rt} covering the full duration of our simulation, to illustrate the periodicity of the burst cycles.

\begin{figure}
	\centering
	\includegraphics[width=\columnwidth]{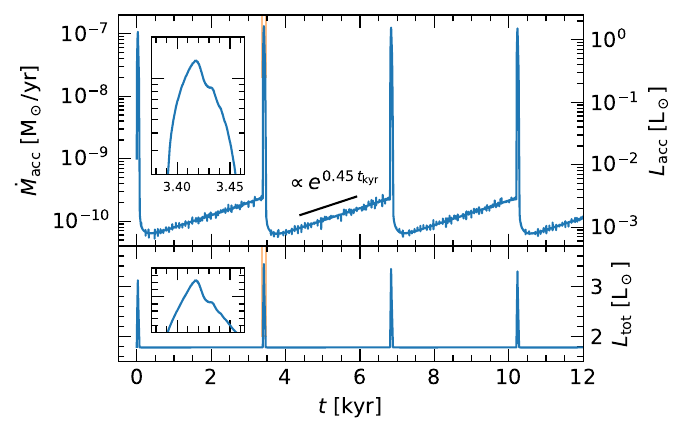}
	\caption{Top: accretion rate onto the star (through the inner radial boundary) and accretion luminosity as functions of time for our fiducial model. The inset zooms in on the narrow burst region highlighted in orange. The black line corresponds to exponential growth with an e-folding time of $\sim\!2.2$\,kyr. Bottom: total stellar luminosity as a function of time.}
	\label{fig:fiducial-mdot-lacc}
\end{figure}

\begin{figure}
	\centering
	\includegraphics[width=\columnwidth]{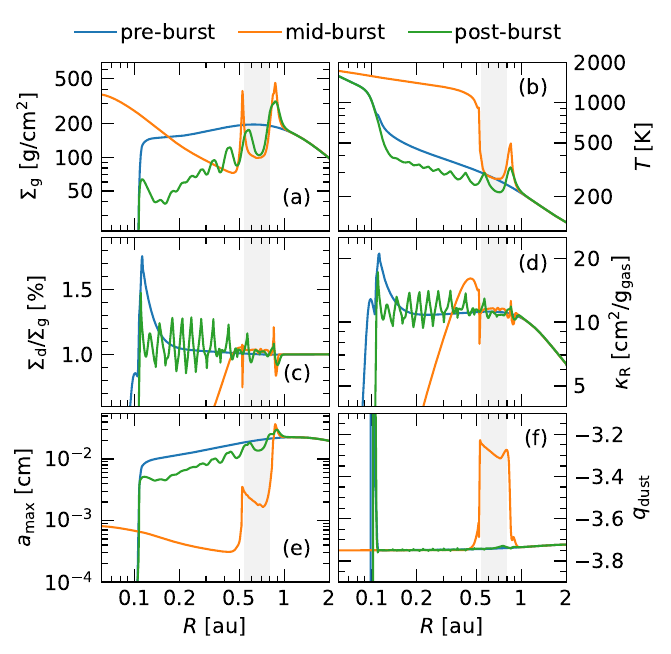}
	\caption{Radial profiles of various gas- and dust-related quantities during the pre-burst (blue), burst (orange), and post-burst (green) phases for our fiducial model: gas surface density (panel \emph{a}), temperature (\emph{b}), dust-to-gas ratio (\emph{c}), Rosseland mean opacity (\emph{d}), maximum grain size (\emph{e}), and dust size distribution exponent (\emph{f}). The gray-shaded region in the burst phase highlights the area where dust is recondensing and recoagulating after the passage of the heating front. The associated movie is available online.}
	\label{fig:fiducial-panels}
\end{figure}

Assuming that all the mass exiting the domain radially inwards is accreted onto the star, we can compute the accretion rate onto the star $\Mdotacc$ as
\begin{equation}
	\label{eq:mdot}
	\Mdotacc = -2\pi R \left[\Sigmag {u_R}_\text{g} + \sum\limits_{i}{\Sigma_{\text{d},i}} {u_R}_{\text{d},i}\right]\Big|_{R=R_\text{in}},
\end{equation}
and the corresponding accretion luminosity $\Lacc$ assuming a stellar radius of $R_\star=2.5\,\text{R}_\odot$ ($T_\star=4200$\,K) as
\begin{equation}
	\label{eq:lacc}
	\Lacc = \frac{GM_\star\Mdotacc}{R_\star}.
\end{equation}
We then plot $\Mdotacc$, $\Lacc$, and the total stellar luminosity $L_\text{tot} = L_\star + \Lacc$ as functions of time in Fig.~\ref{fig:fiducial-mdot-lacc}. We find that both $\Mdotacc$ and $\Lacc$ increase by about three orders of magnitude during the burst phase, reaching peak values of $\sim\!10^{-7}\,\Msun/\text{yr}$ and $\sim\!1\,\Lsun$, respectively, before dropping back to quiescent values of $\sim\!10^{-10}\,\Msun/\text{yr}$ and $\sim\!10^{-3}\,\Lsun$. These values are in good agreement with \cecilt, albeit shifted downwards by a factor of $\sim\!10$ due to the lower reference surface density in our model. The star itself reaches a peak luminosity of $\approx 3.5\,\Lsun$ during the burst phase, with about half of its total luminosity coming from accretion. We note that modulations in the accretion rate due to the reflares can be also seen in the inset on Fig.~\ref{fig:fiducial-mdot-lacc}.

For a more quantitative comparison of the pre- and post-burst states as well as the burst phase itself, we show radial profiles of various gas- and dust-related quantities at different times in Fig.~\ref{fig:fiducial-panels}\footnote{An animation of the radial profiles of $\Sigmag$, $T$, $\Sigmad$, and $\amax$ as a function of time can be found at \url{https://zenodo.org/records/19613864}.}. The disk is practically featureless in the pre-burst state (blue), with the exception of a pile-up of dust near the inner edge of the dead zone at $\approx\!0.12$\,au due to the presence of a pressure maximum induced by the viscosity transition there \citep[see also][]{flock-etal-2017a}. It is nevertheless worth noting that the gas surface density interior to $\sim\!1$\,au is far from a typical power-law profile, instead characterized by a combination of the inherited post-burst profile from the previous cycle and viscous accretion from the outer disk during the quiescent phase. The quiescent phase is terminated with the reignition of the MRI at $\approx\!0.12$\,au, hinted at by a small kink in the temperature profile (top right).

During the burst phase (orange in Fig.~\ref{fig:fiducial-panels}), a heating front propagates out to $\approx0.9$\,au, increasing temperatures above $\TMRI$ and even $\Tsubl$ within that region. This results in both significant dust sublimation due to high temperatures as well as intense fragmentation due to the dramatic increase in $\alpha$ due to the MRI (with $\alphaMRI=0.1$). As a result, the dust-to-gas ratio, maximum grain size, and total opacity plummet within the burst region. As the front recedes due to cooling, dust is allowed to recondense and regrow via coagulation. This process, highlighted in the gray region, is evident by a drop in temperature to $\lesssim500$\,K and the increase in $\qdust$ to values near $-3.2$, indicating a non-equilibrium state dominated by coagulation \pfeilp. A few more reflares occur during the same burst cycle, each propagating outwards less than the previous one (with the second reflare stopping at $\approx0.45$\,au as shown in the figure), until all available mass is accreted and the burst phase is terminated.

The post-burst phase (green in Fig.~\ref{fig:fiducial-panels}) is characterized by an inverted surface density profile within $\sim1$\,au, as predicted by Eq.~\eqref{eq:sigma-post}, albeit rich in radial substructure in both gas and dust, and a cooler temperature profile compared to the pre-burst state. The latter is a consequence of the lower surface densities leading to both less efficient viscous heating ($\propto\Sigmag$) and more efficient radiative cooling ($\propto\Sigmag^{-1}$).

Interestingly, the dust size distribution has already recovered to an equilibrium state dominated by fragmentation with $\qdust\approx-3.75$, suggesting that dust growth occurs on very short timescales, possibly comparable to the burst duration itself. Nevertheless, as we will show in Sect.~\ref{sub:results-tripod-FLD}, dust evolution during the burst phase is actually crucial in determining both the radial extent of the burst region as well as the post-burst state. This is evident by the spike in opacity at the very edge of the heat front during the burst phase (at $R\approx0.45$\,au in panel \emph{d} of Fig.~\ref{fig:fiducial-panels}), which is a direct consequence of dust fragmentation increasing the abundance of small grains with higher opacities. The dust coagulating on timescales comparable to the burst duration is also hinted at by the gradual increase in $\amax$ within the burst region in the bottom panel of Fig.~\ref{fig:fiducial-Rt} (see, e.g., the region between 0.3--0.8\,au at $t\approx$3.4--3.42\,kyr).

Finally, we discuss the ``S-curve'' of the disk at $R=0.3$\,au in Fig.~\ref{fig:fiducial-s-curve} to illustrate the coevolution of the gas surface density and temperature during a burst cycle. The cycle consists of two "stable" branches (highlighted in gray bands in the figure) corresponding to the quiescent, cold state between the end of a burst cycle until the beginning of the next (bottom), and the hot state during which the MRI is active and the disk is rapidly accreting onto the star (top). Starting at the lower branch, and following the blue curve, the disk evolves as follows (see also the numbered points in Fig.~\ref{fig:fiducial-s-curve}):
\begin{enumerate}
	\item The pre-burst state reaches the critical $\Sigmag$ for runaway heating at the DZIE ($\sim\!0.1$\,au).
	\item The activation of the MRI and the resulting increase in $\alpha$ lead to viscous transport away from the triggering region, partly transporting mass outwards, and sequentially igniting the MRI in regions that satisfy $\Sigma\gtrsim\Sigma_\text{g,min}$ (see Eq.~\eqref{eq:sigma-min}), akin to an outward-propagating burst front. This front passes through the highlighted radius (here $R=0.3$\,au) as it advances outwards, resulting in a momentary spike in $\Sigmag$ at this radius, while $T$ continues to rise rapidly.
	\item Temperatures reach $\TMRI$ and quickly exceed $\Tsubl$, preventing further heating due to the thermostat effect. The disk enters a hot state of viscous accretion with $\alpha=\alphaMRI$.
	\item $\Sigmag$ and $T$ have dropped enough for dust to recondense and allow the disk to cool efficiently.
	\item The burst front recedes past this radius while the disk continues to cool.
	\item Additional reflares may trigger at smaller radii (see remaining colors), repeating the process until $\Sigmag$ finally reaches its post-burst value and the disk returns to the quiescent state. While the precise condition for the triggering of such reflares is not entirely clear, it is very likely linked to a balance between viscous and cooling timescales during the evolution of the burst \citep[see also][]{wunsch-etal-2006}.
\end{enumerate}

Overall, the behavior of the burst in terms of gas properties is very similar to that shown in \cecilt, including the presence of multiple reflares, with the addition of a decoupled dust size distribution and its evolution being detailed in this work. The latter shows quite interesting features, with dynamic dust evolution playing a key role in the dynamics of the burst cycle via the modification of the overall opacity due to fragmentation/sublimation, the coagulation of dust on a timescale comparable to that of the burst event itself, and the accumulation of grains on the pressure maxima induced by the burst flares. In the next paragraphs, we will stress on the importance of dust coagulation in capturing the correct burst evolution, as well as the effects of in-plane radiative diffusion.

\subsection{Comparison among thermodynamical prescriptions}
\label{sub:results-tripod-FLD}

In the previous section we showed the evolution of a burst cycle for our fiducial model, where we included dust fragmentation/coagulation but ignored in-plane radiation transport due to its extreme cost in the post-burst state. Here, we will compare that model against one where dust evolution is omitted, and one where we include radiative diffusion through the disk midplane via Eq.~\eqref{eq:source-terms-4} during a burst cycle.

For the model without dust evolution, we prescribe $\qdust=-3.75$ and a radial profile of $\amax$ informed by the results of our fiducial model:
\begin{equation}
	\label{eq:amax-prescribed}
	\begin{split}
	&a_\text{in} = 260\,\mu\text{m}\,\left(\frac{R}{1.26\,\text{au}}\right)^{-0.59},~
	a_\text{out} = 260\,\mu\text{m}\,\left(\frac{R}{1.26\,\text{au}}\right)^{0.38},\\
	&\amax'(R) = a_\text{in}\,a_\text{out} / {\left(a_\text{in}^\zeta + a_\text{out}^\zeta\right)^{1/\zeta}},\quad \zeta = 3.38,\\
	&\amax(R) = \amin + \frac{1}{2}(\amax' - \amin)\left[1 + \tanh\left(\frac{R_\text{au} - 0.11}{0.002}\right)\right].
	\end{split}
\end{equation}
In the above, $a_\text{in}$ and $a_\text{out}$ are power-laws fitted to the inner and outer disk regions, respectively, joined smoothly around $R=1.26$\,au. This profile follows closely the equilibrium profile of $\amax$ in the pre-burst state of our fiducial model (see blue line in bottom panel of Fig.~\ref{fig:fiducial-panels}). The dust fluids otherwise inherit their properties via $\amin$ and $\amax$ in Eq.~\eqref{eq:amax-prescribed} as if \tripod{} were active \citepalias[i.e., effective grain sizes, Stokes numbers, and velocities, see][]{pfeil-etal-2024}.

\begin{figure}
	\centering
	\includegraphics[width=\columnwidth]{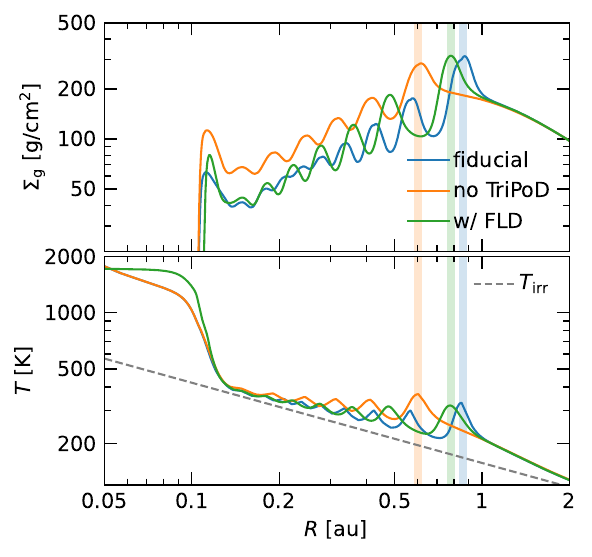}
	\caption{Post-burst gas surface density (top) and temperature (bottom) profiles for models with different thermodynamical prescriptions: our fiducial model with dust evolution but no in-plane radiative diffusion (blue), a model without dust evolution (orange), and a model with in-plane radiative diffusion (green). Vertical lines mark the radial extent of the burst region in each model. A dashed line denotes the irradiation temperature profile.}
	\label{fig:thermo-postburst}
\end{figure}

Both models are initialized very close to the pre-burst state of our fiducial model, at $t=3.37$\,kyr, and evolved through a full burst cycle. The resulting post-burst states in gas surface density and temperature are shown in Fig.~\ref{fig:thermo-postburst}, with the burst regions marked with vertical lines. Both models feature several reflare events and similar post-burst temperature profiles.

The model without dust evolution (orange curves) results in a weaker burst, propagating out to $\approx\!0.6$\,au only, compared to $\approx\!0.9$\,au in the fiducial case (blue), and lasting $\approx60$\,yr compared to $\approx85$ for the fiducial model. This is a direct consequence of the lower opacity in the burst region compared to the fiducial model, due to the lack of fragmentation increasing the abundance of small grains which would otherwise dominate the opacity and promote heating. As a secondary effect, the post-burst state has roughly $1.5\times$ higher surface densities within $\sim1$\,au compared to the fiducial model, as less mass has been accreted onto the star during the weaker burst. This is also reflected in the $\sim1.5\times$ lower accretion rate during the burst (not shown).

As for the model with in-plane radiative diffusion (green curves), the burst region extends to $\approx\!0.8$\,au, slightly less than in the fiducial case. This is likely due to the radial diffusion of heat away from the burst front, resulting in slightly lower peak temperatures. The latter effect can be seen in the post-burst temperature profile (lower panel of Fig.~\ref{fig:thermo-postburst}), which features very slightly smoother temperature peaks compared to the other two models. A by-product of this is an overall slower evolution of the burst, resulting in one less reflare event. Finally, the temperature profile within the always-MRI-active region interior to $\Rrim$ is also different in the model with radiative diffusion.

All in all, we find that the effects of in-plane radiative diffusion are secondary to those of dust evolution during the burst phase, with the former slightly modifying the specifics of the burst evolution but otherwise resulting in a similar post-burst state. In Appendix~\ref{apdx:FLD-postburst}, we further show that in-plane radiative diffusion has no effect on the dynamics during the quiescent, viscous post-burst phase. Dust evolution, on the other hand, plays a key role in determining both the radial extent of the burst as well as the total mass accreted onto the star during a burst event, by a factor of $\sim\!1.5$ in both aspects in our models. For these reasons, we omit in-plane radiative diffusion for the rest of our models in this work, but always include dynamic dust evolution with \tripod{}.

\subsection{Different levels of turbulence in the dead zone}
\label{sub:vary-alpha}

Having analyzed the behavior of our fiducial model in detail and demonstrated the importance of dust evolution, we now shift our focus to models with different values of $\alphaDZ$ to investigate the role of viscous diffusion during and after a burst event. Using the method described in Sect.~\ref{sec:pre-post-constraints}, we can first evaluate whether a model with a given $\alphaDZ$ would contain an unstable configuration in the first place, and find that models with $\alphaDZ\geq 8\,\times10^{-5}$ are indeed prone to accretion bursts for our choice of disk surface density profile and opacity model. We therefore run models with $\alphaDZ=3\times10^{-4}$ and $10^{-4}$ initialized with the pre-burst state computed via Sect.~\ref{sec:pre-post-constraints}, and evolve them through a burst event and several kyr of viscous evolution thereafter. 

\begin{figure}
	\centering
	\includegraphics[width=\columnwidth]{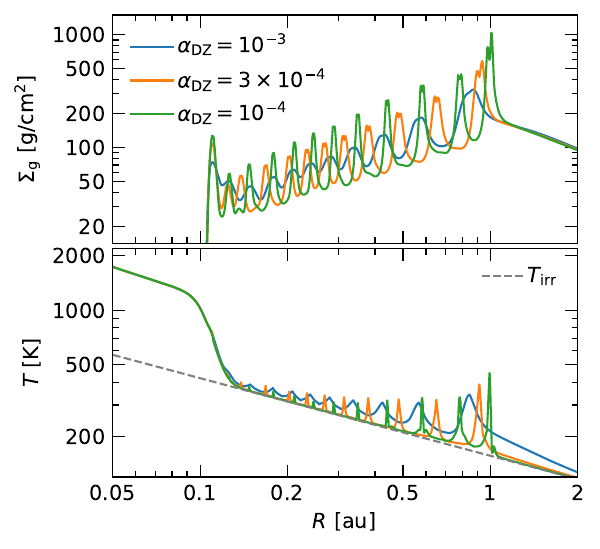}
	\caption{Post-burst states similar to Fig.~\ref{fig:thermo-postburst} for models with different values of $\alphaDZ$. Lower values of $\alphaDZ$ result in more prominent gas substructures due to the less efficient viscous diffusion.}
	\label{fig:alpha-postburst}
\end{figure}

Figure~\ref{fig:alpha-postburst} shows the post-burst gas surface density and temperature profiles for models with different values of $\alphaDZ$. As expected, lower values of $\alphaDZ$ result in more prominent spikes in $\Sigmag$ due to less efficient viscous diffusion during the burst evolution. The weaker viscous dissipation is also reflected in the lower temperatures for lower $\alphaDZ$ values as well, albeit with sharper peaks due to the very high opacities reached at pressure maxima formed during the burst. The burst evolution is otherwise very similar among the different models, which is also expected as its dynamics are mostly governed by the value of $\alphaMRI$ (see Sect.~\ref{sec:pre-post-constraints}). As a result, the burst region extends to approximately the same radius of $\sim\!1$\,au in all models, albeit with a weak inverse scaling of the size of the burst region with $\alphaDZ$, as the surface density in the pre-burst state is higher for lower $\alphaDZ$ (see also Fig.~\ref{fig:pre-post-examples}).

\begin{figure}
	\centering
	\includegraphics[width=\columnwidth]{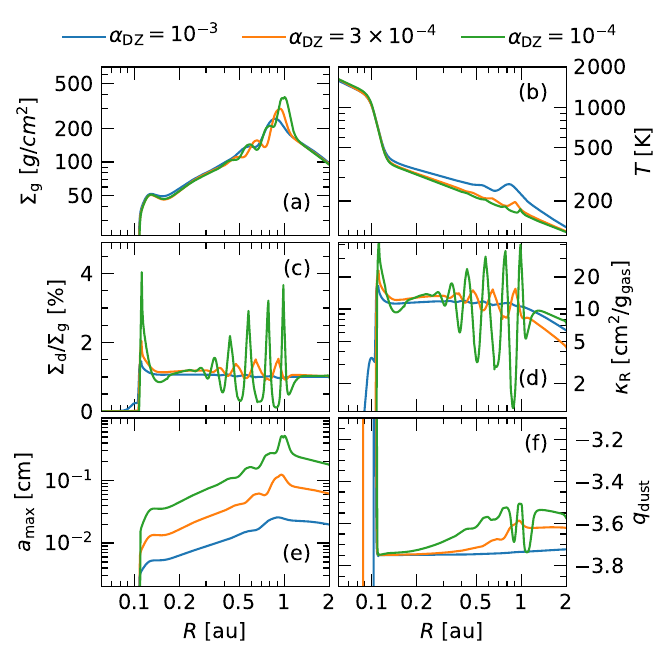}
	\caption{Post-burst states similar to Fig.~\ref{fig:thermo-postburst} for models with different values of $\alphaDZ$. While the gas surface densities are similar among the different models, the effect of lower $\alphaDZ$ in promoting dust growth and accumulation at pressure maxima is evident in panels \emph{c}, \emph{e}, and \emph{f}.}
	\label{fig:alpha-panels}
\end{figure}

However, lower values of $\alphaDZ$ result in significantly longer viscous timescales (i.e., quiescent phase durations), weaker viscous diffusion across the substructures formed during a burst event, and lower turbulent velocities between dust grains. As a result, grains grow to larger sizes for lower $\alphaDZ$, can accumulate more efficiently on the less diffuse pressure maxima, and have significantly longer to pile up before the next burst event can occur. This is illustrated in Fig.~\ref{fig:alpha-panels}, where we plot the same radial quantities as in Fig.~\ref{fig:fiducial-panels} for models with different $\alphaDZ$ values after $\sim 200/\alphaDZ$\,yr of viscous evolution in the post-burst state.

It is worth noting that the dust-to-gas ratio (panel \emph{c}) is notably enhanced at pressure maxima for lower $\alphaDZ$ values, reaching values of $\epsilon\approx0.03$ for $\alphaDZ=10^{-4}$ compared to $\epsilon\approx0.015$ for $\alphaDZ=3\times10^{-4}$. Furthermore, for the same $\alphaDZ$, the dust grows to sizes of $\sim\!3$\,mm, transitioning from small- to intermediate-scale turbulence, evident by the increase in $\qdust$ from $\approx-3.75$ to $\approx-3.5$ (panel \emph{f}). The presence of larger grains and their efficient accumulation at long-lived pressure maxima could enable planetesimal formation via the streaming instability \citep{youdin-goodman-2005,johansen-youdin-2007} in these regions during the quiescent phase, an aspect we will explore in more detail in Sect.~\ref{sec:results-planet-formation}.

\section{Results: planetesimal formation in the quiescent phase}
\label{sec:results-planet-formation}

\begin{figure}
    \centering
    \includegraphics[width=\columnwidth]{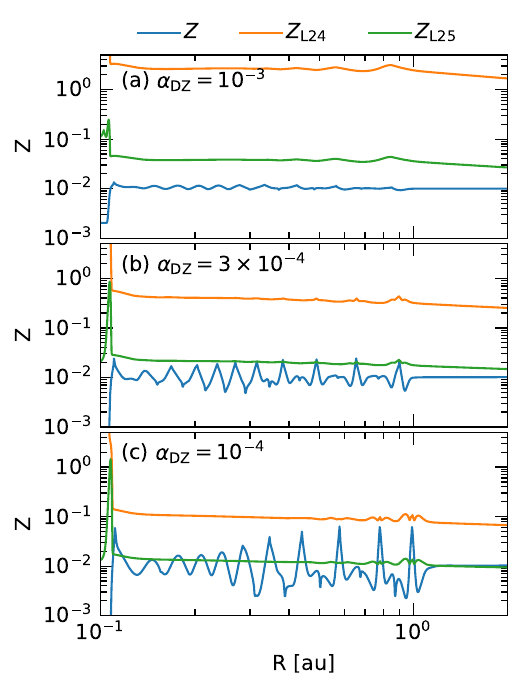}
    \caption{Comparison of the disk metallicity with clumping criteria from Eq.~\eqref{eq:L24} and Eq.~\eqref{eq:L25} after burst for the different viscosity values.}
    \label{fig:Z_crit}
\end{figure}

In this section, we investigate if the dust concentration resulting from the outbursts is high enough to trigger the formation of planetesimals via the streaming instability (SI). Understanding the threshold for the SI to be active (clumping criterion), that is, quantifying the minimum dust concentration $Z = \Sigmad/\Sigmag$ and dynamic pebble size ($\rm St$) necessary to trigger the SI is still an active area of research \citep{youdin-goodman-2005,youdin-johansen-2007,johansen-youdin-2007,Li&Y2021}. To investigate the formation of planetesimals in the quiescent phase, we consider two clumping criteria found in recent literature, namely the ones found in \citet{Lim2024} and \citet{Lim2025} (henceforth ``L24'' and ``L25'', respectively). The first one is given by,
\begin{equation}
\begin{split}
	\label{eq:L24}
    \log Z_\text{L24}(\St_\text{SI}, \alpha) 
    &= 0.15 (\log \alpha)^2 - 0.24 \log \St_\text{SI} \log \alpha \\
    &- 1.48 \log \St_\text{SI} + 1.18 \log \alpha
\end{split}
\end{equation}
for $(\St,\alpha) \in ([10^{-2},10^{-1}],[10^{-4},10^{-3}])$. The second	 is given by,
\begin{equation}
	\label{eq:L25}
    \log Z_\text{L25}(\St_{\rm SI}) = 0.10 (\log \St_\text{SI})^2
    + 0.07 \log \St_{\rm SI} - 2.36
\end{equation}
and was derived for $\St \in [10^{-3}, 1]$ and in the absence of forced turbulence. Since these criteria are derived for a single dust size, in the following section, we assume $\St_{\rm SI} = \St(a_{\rm max})$. The planetesimal formation rate can then be calculated as \citep{Miller2021},
\begin{equation}\label{eq:dSigplan}
    \frac{\text{d}\Sigma_\text{plan}^i}{\text{d}t} = - \zeta \mathcal{P} \Sigmad^i \St^i\Omega_K 
\end{equation}
where $\zeta=0.1$ is the planetesimal formation efficiency per settling timescale and $\mathcal{P}$ is the activation function \citep{Miller2021} given by:
\begin{equation}
    \mathcal{P} = \frac{1}{2} \left[ 1 + \tanh\left( \frac{\log(Z/Z_\text{crit})}{n}\right) \right],\quad Z_\text{crit}\in\{Z_\text{L24},Z_\text{L25}\},
\end{equation}
where $n = 0.03$ is a smoothing factor. We compare the aforementioned clumping criteria with the dust concentration in our setups \SI{100}{yr} after a burst, which can be seen in Fig.~\ref{fig:Z_crit}.

\begin{figure}
    \centering
    \includegraphics[width=\columnwidth]{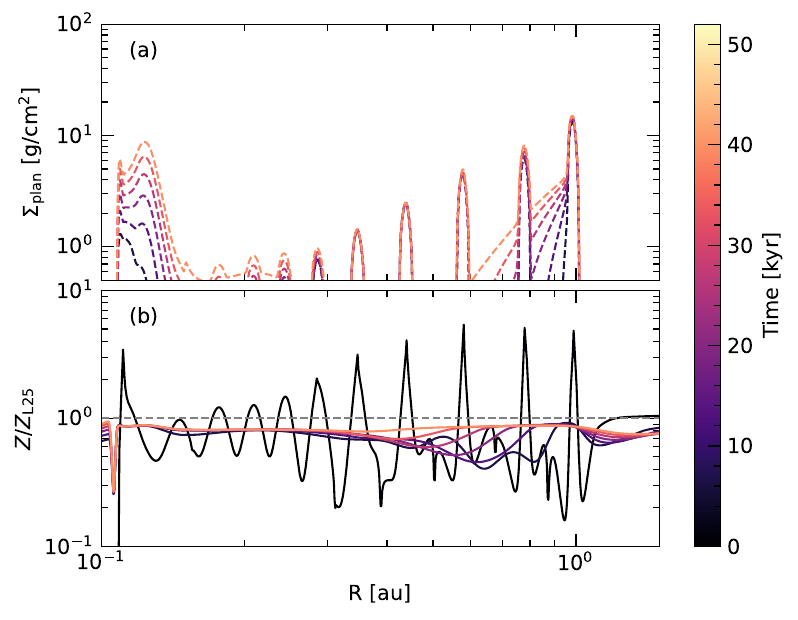}
    \caption{The planetesimal surface density (panel \emph{a}) and the ratio of disk metallicity to critical metallicity (panel \emph{b}) throughout the \texttt{TriPoDPy} simulation}
    \label{fig:alpha1e-4_formation}
\end{figure}

As we can clearly see, the only simulation/criterion pairing that predicts any significant amount of planetesimal formation is the $\alpha = 10^{-4}$ setup with the L25 threshold. To actually model the formation of planetesimals in this case, we evolved the disk from the hydrodynamical simulation starting with its after-burst state 700\,yr after the end of the burst with an isothermal 1D simulation where we assumed a fixed $T(R)$ (and therefore $\alpha(R)$) for the duration of the simulation. This approach holds well for a large fraction of the quiescent phase. We used the \texttt{TriPoDPy} code \citep{Kaufmann2025}, which includes viscous gas evolution and uses the \texttt{TriPoD} method for the dust evolution and behaves equivalently to the hydro simulation in the quiescent phase (see Appendix~\ref{apdx:Tripodpy-Hydro-comp} for a comparison), but is significantly faster as it omits both temperature and momentum evolution. The formation of planetesimals is modeled as a sink term given by Eq.~\eqref{eq:dSigplan} with the L25 clumping criterion. The resulting planetesimal surface density and the ratio of disk and clumping metallicity can be seen in Fig.~\ref{fig:alpha1e-4_formation}.

After evolving the system for $50$\,kyr, we form $1.6\,\Mearth$ in planetesimals, which is split between the fast planetesimal formation at the pileups from the burst and the continuous formation at the inner rim. This is also neatly illustrated by the fact that the SI criterion is significantly surpassed in the bumps at the beginning, and the continuous formation of planetesimals keeps $Z/Z_{\rm crit} \approx 1$ at the inner edge.

We have to note that this formation of planetesimals triggered by outbursts is a very tentative result. Firstly, the criteria for SI were only met for one of the parameter setups ($\alpha=10^{-4}$), considering the most favorable SI criterion. Additionally, there remains a large uncertainty with regard to the appropriate clumping criteria to use, as both the parameter space and setup where they were derived do not fully cover our simulations (e.g., considering the presence of multiple dust sizes). Also, connecting the SI being active to a planetesimal formation rate is still uncertain, with most models, including this work, assuming a constant formation rate per settling timescale \citep{Miller2021,Schoonenberg2018,Lau2022}.

% \section{Results: synthetic observations wth ngVLA}
% \label{sec:results-ngvla-synthetic-observations}

% \alexnote{Sample}

\section{Discussion}
\label{sec:discussion}

Here we discuss the implications of our results in the context of accretion outbursts and planet formation, as well as the limitations of our models. We also compare and contrast our models with more typical FU Orionis models in the literature.

\subsection{Comparison to ``traditional'' FU Ori models}
\label{sub:discussion-comparison-fuori}

FU Orionis-type outbursts have been modeled extensively in the literature, with a major difference being the accretion rate and luminosity during the burst phase. In particular, typical FU Ori models feature peak accretion rates of $\sim\!10^{-4}$--$10^{-3}\,\Msun/\text{yr}$ \citep[e.g.,][]{zhu-etal-2009,zhu-etal-2010,bae-etal-2013,bae-etal-2014,vorobyov-basu-2015}, resulting in accretion luminosities of $\sim\!10^2$--$10^3\,\Lsun$ for solar-type stars. These values are significantly higher than those found in our models (see Fig.~\ref{fig:fiducial-mdot-lacc}), and is mainly due to our choice of a much lower surface density profile. In fact, the aforementioned models leverage the high surface densities to drive the gravitational instability throughout the disk, which in turn supplies mass to the inner disk with an $\alphaDZ$ that can be as high as $10^{-2}$ \citep[e.g.,][]{zhu-etal-2009,vorobyov-basu-2015}, resulting in much higher pre-burst accretion rates and more massive burst events that can extend out to several au (see also dashed line in Fig.~\ref{fig:pre-post-examples}).

In fact, considering both the period and amplitude of the accretion bursts in our models (every $\sim$kyr, with $\Lacc\sim3{\Lacc}_0$), our fiducial model results are unlike those of typical stellar outbursts seen in YSOs \citep[see Fig.~3 in][]{fischer-etal-2023}, but can be easily tuned to match with those of EX Lupi-type objects \citep{herbig-2007} or even FU-Ori-like objects by adjusting the disk mass and the values of $\alphaDZ$ and $\alphaMRI$ accordingly. An exact match with such objects is, however, not the focus of this work.

\subsection{Constraints on turbulent dissipation and their implications}
\label{sub:discussion-constraints-turbulence}

As we have seen both in our analytical estimates in Sect.~\ref{sec:pre-post-constraints} and in our numerical models in Sect.~\ref{sec:results-outbursts}, the values of both $\alphaDZ$ and $\alphaMRI$ play key roles in determining the characteristics of accretion outbursts near the DZIE by setting, the pre-burst state, the extent of the burst region, and assigning a lifetime to both the quiescent phase and the radial substructures formed during the burst phase. In this work, we have adopted fiducial values of $\alphaDZ$ and $\alphaMRI$ following \cecilt, without actually arguing for the origin of a turbulent mechanism that would yield these values.

In particular, while the value of $\alphaMRI=0.1$ we have adopted is well within the range of turbulent stress reported for the MRI from both ideal and non-ideal MHD models in the literature \citep[e.g.,][]{simon-etal-2011,bai-stone-2013a,flock-etal-2017a,iwasaki-etal-2024}, the origin of a non-zero $\alphaDZ$ in the dead zone is far less clear. Mechanisms that are often invoked in the outer disk such as the vertical shear instability \citep[VSI,][]{nelson-etal-2013,flock-etal-2020b}, an ambipolar-diffusion-mediated MRI \citep[e.g.,][]{turner-etal-2014,delage-etal-2022}, or even the gravitational instability \citep[GI,][]{toomre-1964,gammie-2001} are not expected to operate efficiently within the inner 10\,au of the disk due to prohibitively long cooling timescales \citep{lin-youdin-2015}, low ionization fractions \citep{bai-stone-2013b}, and high values of the Toomre parameter $Q$ \citep{toomre-1964}, respectively.

Possible sources of turbulence within a few au from the central star could include hydrodynamical instabilities such as the convective overstability \citep[COS,][]{klahr-hubbard-2014,lyra-2014} or the zombie vortex instability \citep[ZVI,][]{marcus-etal-2015,marcus-etal-2016}, both of which, however, are very sensitive to the disk's thermal relaxation timescale, vertical stratification, and radial density and temperature profiles. As a result, the efficiency of these instabilities in driving adequate turbulence in the dead zone remains to be confirmed with high-resolution non-linear simulations including realistic thermodynamics. Finally, the turbulence driven by the streaming instability itself is expected to be very weak, with $\alpha\lesssim10^{-5}$ \citep{baronett-etal-2024}.

Nevertheless, assuming that the bulk of the accretion is driven via a magnetothermal wind at the disk surface layers \citep{bai-stone-2013b}, all that is really required is some level of turbulent diffusion at the DZIE to enable the ``ignition'' of the MRI via viscous heating. To that end, \citet{jankovic-etal-2021} and \citet{iwasaki-etal-2024} have shown that the transition from the MRI-active inner disk to the dead zone may be mediated by a radial buffer zone with a nonzero $\alpha$, which could potentially play the role of the ``spark'' that ignites the burst cycle. This could be modeled with a combination of imposing a laminar radial flow \citep[e.g.,][]{kimmig-etal-2020} and a different transition region between the active and dead zones via, for instance, a more detailed temperature dependence of $\alpha$ in Eq.~\eqref{eq:alpha-mri} informed by the aforementioned works.

Although modeling the triggering of the MRI via a temperature threshold, $T_\mathrm{MRI}$, is commonly used in simulations of outburst mechanisms \citep[e.g.,][]{Zhu2009, bae-etal-2013, Kadam2020, Steiner2021, cecil-flock-2024}, a more detailed determination of MRI-active and dead zones can be achieved by coupling the MRI activity to the ionization fraction and non-ideal MHD effects in the inner disk \citep{Dzyurkevich2013, Mohanty2018, jankovic-etal-2021, delage-etal-2022}. This can lead to a different position and shape of the DZIE, which then additionally depends on the structure and strength of the magnetic fields interacting with the disk material. Recent studies of steady-state solutions for large-scale magnetic fields in the inner disk \citep[e.g.,][]{Steiner2025} in combination with tabulated diffusivities of non-ideal MHD effects \citep[e.g.,][]{Desch2015, Williams2025} will be used in a forthcoming paper to investigate the influence of a more elaborate MRI activity description on the triggering and evolution of the outburst mechanism analyzed in this work.

It is worth noting that embedded planets will contribute to a significant fraction of the angular momentum transport within the inner disk via the excitation of spiral density waves \citep{goldreich-tremaine-1979,ogilvie-lubow-2002}, effectively driving a non-zero $\alpha$ \citep{goodman-rafikov-2001}. In a low-viscosity environment such as the dead zone, such planets could even open deep gaps whose edges will be prone to the RWI \citep{ziampras-etal-2025b}, further enhancing the total Reynolds stress within the dead zone. Finally, shock heating by such planets can act as an additional, very efficient source of thermal energy especially in the inner few au of the disk \citep{rafikov-2016,ziampras-etal-2020a,rowther-etal-2020,ono-etal-2025, okuzumi-etal-2025}, which has the potential to eliminate the need for viscous heating altogether to trigger the MRI at the DZIE. The presence of giant, gap-opening planets in the inner disk could also lead to accumulation of gas near the DZIE, possibly leading to burst events of similar origin but different behavior to those studied here \citep{lodato-clarke-2004}. Investigating these possibilities could be the focus of future work.

\subsection{Implications for planet migration}
\label{sub:discussion-planet-migration}

The inverted or otherwise heavily modified surface density profiles formed during a burst event and maintained during the majority of the quiescent phase (see top panels of Fig.~\ref{fig:pre-post-examples}), the presence of multiple pressure maxima for several kyr after a burst (see Fig.~\ref{fig:alpha-postburst}), and even the non-power-law-like temperature profiles for sufficiently high $\alphaDZ$ (bottom panels of Fig.~\ref{fig:pre-post-examples}), could have significant implications for the migration of low-mass planets embedded in the inner disk. If indeed there is sufficient turbulent diffusion to prevent planets from opening gaps in the gas \citep[e.g.,][]{crida-etal-2006,duffell-2015,ziampras-etal-2025b}, then such planets will be subject to type-I migration \citep{goldreich-tremaine-1980,ward-1997a}, a process known to be highly sensitive to the local radial gradients of both surface density and temperature \citep{paardekooper-etal-2010,paardekooper-etal-2011}.

In this context, pressure bumps could act as planet traps \citep[e.g.,][]{coleman-nelson-2014,izidoro-etal-2017}, halting inward migration and promoting the growth of planetary embryos via pebble accretion \citep{lambrechts-johansen-2012,lambrechts-etal-2014}. Once planets have grown sufficiently massive to no longer migrate in the type-I regime, the DZIE itself would then act as a barrier, preventing further inward migration \citep{ataiee-kley-2021,chrenko-etal-2022}. Our analytical prescriptions for the pre- and post-burst states from Sect.~\ref{sec:pre-post-constraints} can be used to model planet migration in the inner regions of protoplanetary disks qualitatively, without the need for full hydrodynamical simulations.

We stress that these expectations hinge on the assumption that the burst evolution is not strongly affected by an embedded planet, which is not straightforward to assume given that planetary spirals could either open gaps that can modify the local cooling properties of the disk or smear the burst-related pressure bumps, with either process affecting the evolution of the burst. The interplay between accretion outbursts and embedded planets will be the subject of future work.

\subsection{Limitations of a 1D model}
\label{sub:discussion-limitations-1d}

Our models are focused on the radial evolution of the inner regions of the protoplanetary disk during and after an accretion outburst, and as such are limited to a vertically integrated, axisymmetric (1D) framework. While this approach allows us to capture the key physical processes involved in the burst dynamics and dust evolution, it comes with some simplifying assumptions and limitations to what we can infer from our results.

Possibly the most significant limitation of our 1D approach is the inability to capture nonaxisymmetric instabilities that may arise during the burst phase, in particular the Rossby wave instability \citep[RWI,][]{lovelace-1999}. The RWI is known to develop where the vortensity profile of the disk $\varpi(R) = (\nabla\times\vel)/\Sigma\cdot\hat{z}$ features a local extremum \citep[see also][]{chang-youdin-2024}, which is easily satisfied around the burst front even during the evolution of the outburst \citep[see also][]{cecil-flock-2024,cecil-etal-2026}.
Effectively, the development of the RWI would erode the burst front into one or more vortices, weakening its sharpness and possibly the extent to which dust grains can accumulate at the resulting pressure maximum. The Reynolds stress generated by the vortices may also contribute to the overall angular momentum transport during the burst phase and while the vortices persist \citep{kuznetsova-etal-2022}, which can last several thousand orbits at $\alpha=10^{-4}$ \citep[e.g.,][]{rometsch-etal-2021}, effectively increasing the value of $\alphaDZ$ during that time and leading to faster turbulent diffusion of substructures formed during the burst. At the same time, however, the presence of vortices would promote dust accumulation by creating long-lived, nonaxisymmetric dust traps, possibly enhancing planetesimal formation via the streaming instability \citep[e.g.,][]{raettig-etal-2015}. The interplay between the RWI, burst dynamics, and dust evolution will be explored in followup work (Ziampras et al., in prep.).

Regarding the vertical structure of the disk, our 1D models assume vertical hydrostatic equilibrium and a vertically isothermal temperature profile, that is, ignoring the hot corona of the disk atmosphere. We expect that this is a reasonable approximation for both the temperature structure and the gas--dust coevolution near the midplane, as that is where most of the mass is found and where viscous heating dominates the thermal budget, especially during the burst phase. However, we note that the 2D axisymmetric models of \cecilt{} have shown vertical stirring of gas during the burst phase, which may affect the vertical distribution of dust grains for a brief period of time. The implications of this effect on dust evolution and planetesimal formation remain to be explored in future work.

Finally, it is entirely possible that the assumption of a smooth, power-law-like profile for the flaring angle $\theta(R)$ (see Eq.~\eqref{eq:theta-rim}) in the outer disk may not fully hold in reality, due to the presence of shadows cast by the puffed-up, directly illuminated inner rim \citep[see also][]{dullemond-monnier-2010,flock-etal-2025}. Such shadows could simply modify the irradiation heating term $\Qirr$ in Eq.~\eqref{eq:navier-stokes-2}, or even trigger additional thermal instabilities in the outer disk \citep{wu-lithwick-2021,melon-fuksman-klahr-2022,sudarshan-etal-2025}. Nevertheless, given the relatively small radial extent of interest in our models ($\lesssim5$\,au), and the central role of viscosity in the burst evolution, we expect that the effects of radial modulations to the irradiation heating would be secondary to those of viscous heating and cooling.

\newpage

\section{Summary}
\label{sec:summary}

In this work we have presented vertically integrated, axisymmetric models of stellar outbursts via a thermal instability at the dead zone inner edge, including for the first time a fully integrated treatment of dust coagulation, dust--gas thermal and dynamical coupling, and radiative transfer via surface cooling, in-plane radiative diffusion, viscous heating, and stellar irradiation. While our models are limited to a 1D framework, several key messages emerge from our analysis.

We have developed a semi-analytical method that can be used to compute the pre- and post-burst states of a disk subject to accretion outbursts at the DZIE for a given set of stellar and disk parameters. We then demonstrated that our method yields results in excellent agreement with numerical simulations employing full radiation hydrodynamics and dust evolution. This method can be used to efficiently scan the parameter space of accretion outbursts without the need for expensive numerical simulations.

Dust evolution plays a key role in determining the radial extent of the burst region during an accretion outburst---moreso than radiative diffusion---due to dust fragmentation increasing the abundance of small, opacity-carrying grains. As a result, our models including dust evolution yield up to 1.5$\times$ larger burst regions and accreted mass onto the star compared to models without dust evolution, both of which have a noticeable effect on the burst amplitude and post-burst density distributions.

The level of turbulent diffusion in the dead zone influences dust growth and planetesimal formation in multiple ways. Regarding the gas, lower values of $\alphaDZ$ result in more massive bursts that extend further out in radius, more pronounced and longer-lasting substructures due to the weaker viscous diffusion, and much longer quiescent periods, all of which act in favor of more efficient and longer-lived dust traps. As for the dust, lower $\alphaDZ$ values lead to lower turbulent velocities and as a result larger grains, which can accumulate more efficiently at pressure maxima, all while further promoting trapping by reducing radial diffusion of dust grains.

Even though the outbursts lead to an accumulation of dust in several rings, it remains questionable whether the SI is triggered in these bumps. In our follow-up simulations, we found that only the most favorable SI criterion (L25) and $\alpha = 10^{-4}$ formed any planetesimals.

While our models combine several intertwined physical processes in a self-consistent manner, the limits of a 1D framework prevent us from capturing nonaxisymmetric instabilities such as the Rossby wave instability, which is expected to develop at the burst front during the outburst phase \citep{cecil-etal-2026}. The implications of such instabilities on the burst dynamics and dust evolution will be explored in followup work.

% \newpage
\begin{acknowledgements}
AZ would like to thank Mario Flock and Zhaohuan Zhu for insightful discussions.
AZ, TB, and NK acknowledge funding from the European Union under the European Union's Horizon Europe Research and Innovation Programme 101124282 (EARLYBIRD). Views and opinions expressed are those of the authors only. TB acknowledges funding by the Deutsche Forschungsgemeinschaft (DFG, German Research Foundation) under Germany's Excellence Strategy - EXC-2094 - 390783311. This research was supported in part by grant NSF PHY-2309135 to the Kavli Institute for Theoretical Physics (KITP). All plots in this paper were made with the Python library \texttt{matplotlib} \citep{hunter-2007}. %Typesetting was expedited with the use of GitHub Copilot, but without the use of AI-generated text.
\end{acknowledgements}

\section*{Data Availability}

Data from our numerical models are available upon reasonable request to the corresponding author.

\bibliographystyle{aa}
\bibliography{refs}

% \clearpage
% \newpage

\appendix

\section{Comparison between \pluto{} and Sect.~\ref{sec:pre-post-constraints}}
\label{apdx:pre-post-burst}

Here we compare the pre- and post-burst states obtained with the method described in Sect.~\ref{sec:pre-post-constraints} against our fiducial model from Sect.~\ref{sub:results-fiducial-model}, which includes full radiation hydrodynamics and dust evolution. For this comparison, the dust opacity includes a snapshot of the dust distribution from the fiducial model at $t=3.37$\,kyr, just before the burst event begins. We therefore set $\amax$ via Eq.~\eqref{eq:amax-prescribed}, $\qdust=-3.75$, and a constant dust-to-gas ratio $\epsilon=0.01$. The latter two approximations hold very well during the viscous quiescent phase (see Fig.~\ref{fig:alpha-panels}, blue curves).

\begin{figure}
	\centering
	\includegraphics[width=\columnwidth]{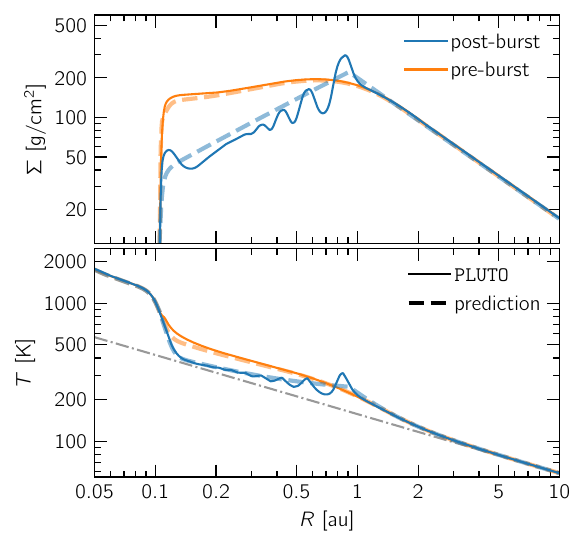}
	\caption{Comparison of pre- and post-burst states in gas surface density (top) and temperature (bottom) obtained with \pluto{} (solid lines) and with the method described in Sect.~\ref{sec:pre-post-constraints} (dashed lines). The two methods show excellent agreement overall.}
	\label{fig:pre-post-comparison}
\end{figure}

Figure~\ref{fig:pre-post-comparison} summarizes our findings, showing the pre- and post-burst states obtained with \pluto{} (solid lines) and with the method outlined in Sect.~\ref{sec:pre-post-constraints} (dashed lines). We find excellent agreement in the overall behavior, with key differences being the presence of substructures in the \pluto{} results due to the dynamic evolution of the burst, and slightly lower density/temperatures at the inner edge of the dead zone when using our method. The latter is most likely due to the change in $\amax$ as the disk evolves viscously, a process not captured by our method, where $\amax$ is fixed to the pre-burst profile. Nevertheless, this effect is minor, and does not affect the overall validity of our approach.

\section{In-plane radiative diffusion during the quiescent phase}
\label{apdx:FLD-postburst}

\begin{figure}
	\centering
	\includegraphics[width=\columnwidth]{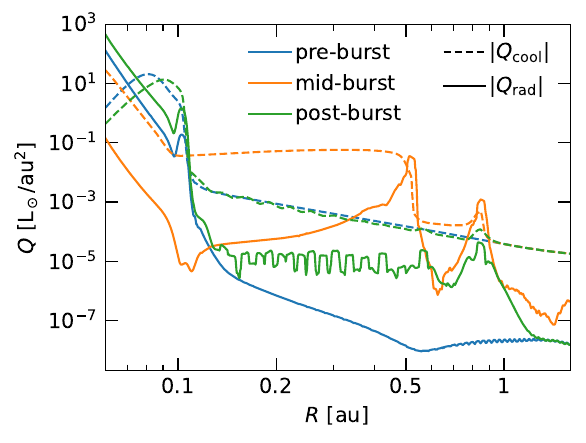}
	\caption{Comparison of the cooling terms $\Qrad$ (solid lines) and $\Qcool$ (dashed lines) during different stages of the burst cycle for our fiducial model. While in-plane diffusion is comparable to thermal cooling around the burst front during the burst phase (orange), it is negligible otherwise.}
	\label{fig:cool-vs-rad}
\end{figure}

To justify our choice of omitting in-plane radiative diffusion during the post-burst, quiescent phase in our models, we compared the cooling terms $\Qcool$ and $\Qrad$ (see Eqs.~\eqref{eq:source-terms-3}~and~\eqref{eq:source-terms-4}) during different stages of the burst cycle in our fiducial model. The results are summarized in Fig.~\ref{fig:cool-vs-rad}, where we plot the two terms during the pre-burst, mid-burst, and post-burst phases.

We find that in-plane radiative diffusion is only relevant near the burst fronts formed during the burst phase, exceeding thermal cooling by a factor of 10 or even 100 near the two fronts at $R\sim\!0.9$\,au and $\sim\!0.5$\,au, respectively. However, $\Qrad$ is otherwise negligible compared to $\Qcool$ during the quiescent pre- and post-burst phases, with the exception of the outermost bump at $\sim \!0.9$\,au post-burst, where $\Qrad\sim0.3\,\Qcool$. This comparison highlights that in-plane radiative diffusion does have an effect on the disk temperature profile and the propagation of the burst fronts during the burst phase (see also Fig.~\ref{fig:thermo-postburst}), but also that our choice to omit it during the quiescent phase is well justified.

\section{\texttt{growpacity}: a computationally efficient dust opacity model suitable for coagulation models}
\label{apdx:growpacity}

\texttt{growpacity} is a Python module designed to provide temperature-dependent Rosseland and Planck mean opacities for a dynamically evolving dust size distribution, particularly suited for dust coagulation models such as \tripod{}. To do so, it provides an interface to \optool{} \citep{dominik-etal-2021}, a tool that can compute accurate opacities for a given dust composition and size distribution, and then precomputes a grid of mean opacities as functions of temperature $T$, maximum grain size $\amax$, and size distribution exponent $\qdust$. These opacity tables are quite lightweight (typically $\lesssim1$\,MB), and can be efficiently interpolated over during a hydrodynamical simulation to provide up-to-date opacities as the dust size distribution evolves.

\begin{figure}
	\centering
	\includegraphics[width=\columnwidth]{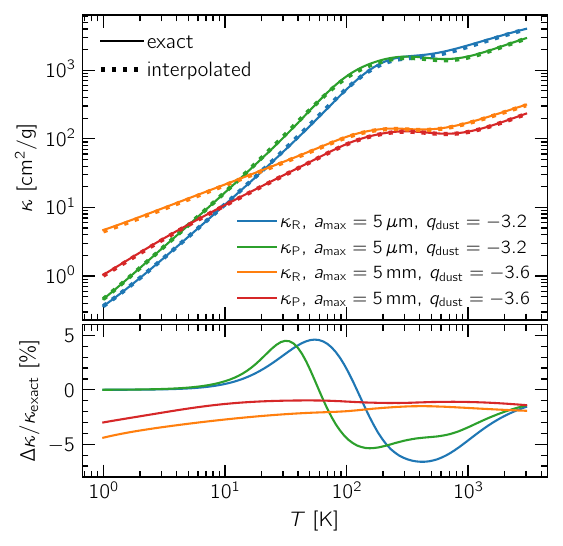}
	\caption{Comparison between exact calculations with \optool{} and their interpolated counterparts for two different combinations of $\amax$ and $\qdust$, representing a growing (fully grown) distribution with $\amax=5\,\mu$m ($\amax=5$\,mm) and $\qdust=-3.2$ ($\qdust=-3.6$). These values have been chosen to lie as far as possible from the sampled grid points, to best test the interpolation scheme. Overall, the interpolated opacities agree very well with the exact calculations, with deviations of $<7\%$.}
	\label{fig:growpacity-verification}
\end{figure}

For a given grain composition and assuming a grain size distribution characterized by $\amin$, $\amax$, and $\qdust$ (see Eq.~\eqref{eq:qdust}), the \optool{} package \citep{dominik-etal-2021} can compute the absorption and scattering opacities $\kabs(\nu)$ and $\ksca(\nu)$ (in cm$^2$/g) as well as the asymmetry factor $g(\nu)$ \citep{henyey-greenstein-1941} over a frequency grid $\nu$. This calculation is done by default using the Distribution of Hollow Spheres method \citep[DHS,][]{min-etal-2005}. The Rosseland and Planck mean opacities $\kappaR$ and $\kappaP$ can then be computed as
\begin{equation}
    \label{eq:kappaRP}
    \begin{split}
	\kappaR(T) &= \frac{\int_0^\infty u_\nu(T) d\nu}{\int_0^\infty [\kabs+(1-g)\,\ksca]^{-1} u_\nu(T) d\nu}, \quad u_\nu(T) = \left.\frac{dB_\nu}{dT}\right|_T, \\
	\kappaP(T) &= \frac{\int_0^\infty \kabs B_\nu(T) d\nu}{\int_0^\infty B_\nu(T) d\nu},
	\end{split}
\end{equation}
where $B_\nu(T)$ is the Planck function at temperature $T$. By fixing the grain composition and $\amin$, \optool{} can be used to calculate the absorption and scattering opacities over a grid of $\amax$ and $\qdust$. Equation~\ref{eq:kappaRP} can then be used to compute and tabulate $\kappaR$ and $\kappaP$ over $\amax$, $\qdust$, and $T$. We found that sampling $\amax$ with 13 points per decade between $0.1\,\mu$m and $10$\,cm, $\qdust$ with 9 points between $-4.5$ and $-2.5$, and $T$ with 300 points between 1--3000\,K results in accurate mean opacities during interpolation while keeping the size of the opacity tables to merely $\sim300$\,kB. A verification test comparing the mean opacities computed with \optool{} directly against those obtained via interpolation in the precomputed tables is shown in Fig.~\ref{fig:growpacity-verification}, demonstrating very good agreement overall.

To evaluate the opacity for a given $\qdust$, $\amax$, and $T$, we can interpolate within the 3D tables of $\kappaR(\qdust,\amax,T)$ and $\kappaP(\qdust,\amax,T)$. We choose to interpolate for $\log\kappa$ as a function of $\qdust$, $\log\amax$, and $\log T$, with regular sampling in this space (i.e., logarithmic spacing for $\amax$ and $T$). This works best when the mean opacities follow a power-law with respect to temperature $\kappa\propto T^b\Rightarrow \log\kappa\propto b\log T$, which is a reasonable approximation and holds especially well for small grains \citep[e.g.,][]{bell-lin-1994,semenov-etal-2003}. It also ensures that the interpolated opacities are always positive in case extrapolation would be needed, although by default we clamp the input parameters to the bounds of the precomputed tables as a safeguard.

\begin{figure}
	\centering
	\includegraphics[width=\columnwidth]{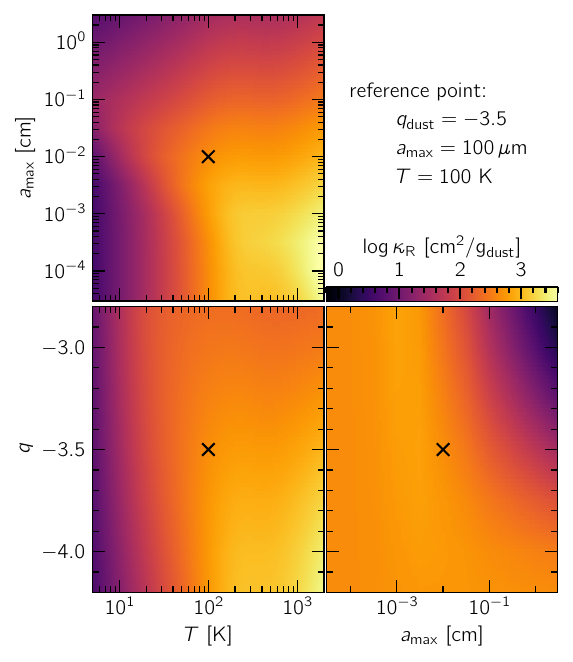}
	\caption{Heatmaps of interpolated Rosseland mean opacities as functions of pairs of parameters, with the third parameter otherwise fixed at a representative value.}
	\label{fig:growpacity-showcase}
\end{figure}

We use a trilinear interpolation scheme, which is fast and simple to implement, and takes advantage of the fact that the arrays $\qdust$, $\log\amax$, and $\log T$ are sorted and regularly spaced to efficiently locate the indices of the grid points that surround the point of interest. For each array $x\in\{\qdust,\log\amax,\log T\}$ and for a target value $x_\text{t}$, we first find the index $i$ such that $x_i\leq x_\text{t} < x_{i+1}$ as $i = \lfloor(x_\text{t}-x_0)\,\Delta x^{-1}\rfloor$, where $x_0$ is the first (smallest) value in the sampling space and $\Delta x = x_{i+1}-x_{i}$ is the (constant) sampling spacing. As this information is known \emph{a priori}, this reduces the complexity of finding the required indices from $\mathcal{O}(\log N)$ to $\mathcal{O}(1)$, where $N$ is the number of grid points in $x$, and is especially efficient for larger grids. The interpolation scheme, implemented in both \texttt{C} and Python, is included in the \texttt{growpacity} package. An example of the interpolated mean opacities as functions of $\qdust$, $\amax$, and $T$ is shown in Fig.~\ref{fig:growpacity-showcase}.

We underscore that \texttt{growpacity} does not provide a new or more realistic dust opacity model, but rather one suitable for use in coagulation models, where dust densities and distributions can vary as a function of position and time---the applicability of the model depends on entirely user-defined choices. The method can be easily extended to include gas opacities \citep[e.g.,][]{semenov-etal-2003,malygin-etal-2014} and prescriptions for the sublimation of dust species \citep[e.g.,][]{isella-natta-2005}, for a more complete opacity model in regimes where gas opacities are significant.

\section{comparing \texttt{TriPoDPy} with \texttt{PLUTO} simulations during the quiescent phase}
\label{apdx:Tripodpy-Hydro-comp}

\begin{figure}
	\centering
	\includegraphics[width=\columnwidth]{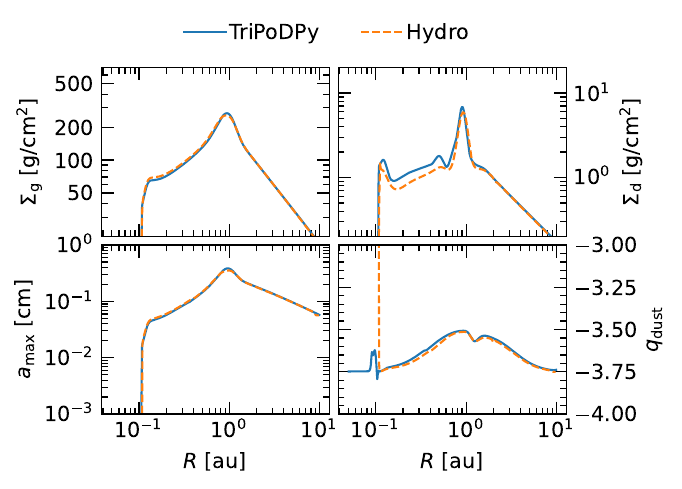}
	\caption{Comparison between the state of the simulation at $t = 20$\,kyr in \texttt{TrPoDPy} and \texttt{PLUTO} showing the gas surface density (top left), the total dust surface density (top right), the maximal grain size (bottom left) and the dust size distribution exponent (bottom right).} 
	\label{fig:tripodpy-comp}
\end{figure}

To ensure that the evolution in the quiescent phase with \texttt{TriPoDPy} matches the full hydro simulations, we ran a test case evolving the $\alpha = 10^{-4}$ simulation from $t= 8$\,kyr to $20$\,kyr without planetesimal formation and compared the final state of the two simulations, which is depicted in Fig. \ref{fig:tripodpy-comp}. As we can see, the gas surface density and the maximal grain size match perfectly, whereas there is a slight difference in the total dust surface density, which can be attributed to the fact that the fluids don't exchange momentum in \texttt{TriPoDPy} and the flux through the MRI-active front is different in the two codes, which is also shown by the dust size distribution exponent in the cavity. For the purposes of investigating the planetesimal formation in the quiescent phase, the simulations behave the same, as the major differences are in the cavity, which does not contribute to the formation of planetesimals in any case.

\section{Full duration radius--time heatmaps for the fiducial model}
\label{apdx:fiducial-Rt-extended}

\begin{figure}[b]
	\centering
	\includegraphics[width=\textwidth]{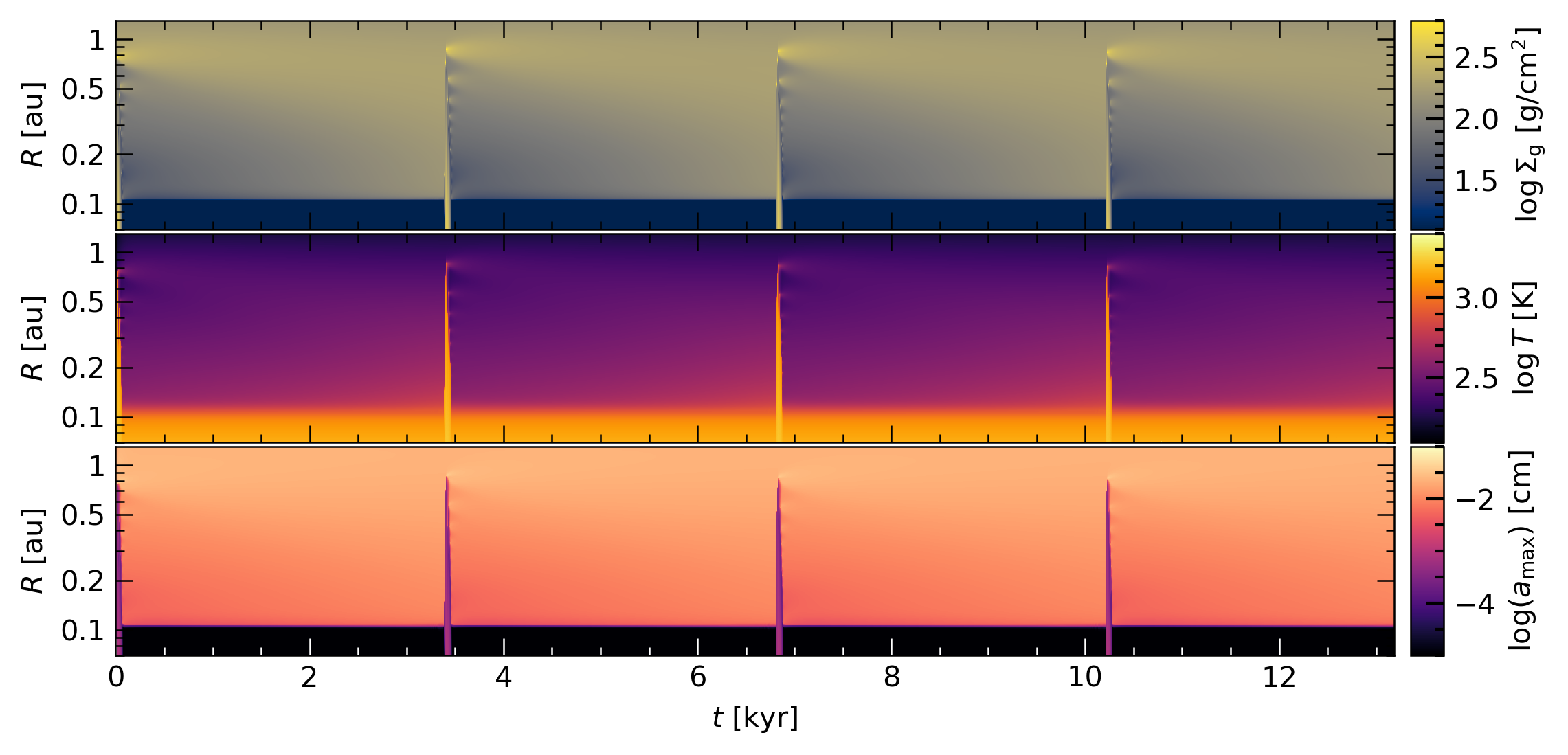}
	\parbox{\textwidth}{\caption{Radius--time heatmaps similar to those in Fig.~\ref{fig:fiducial-Rt}, spanning the entire duration of the simulation.}
	\label{fig:fiducial-Rt-extended}}
\end{figure}

In Fig.~\ref{fig:fiducial-Rt-extended} we show the radius--time heatmaps of gas surface density, temperature, and dust surface density for the entire duration of the simulation of our fiducial model. The figure illustrates the clear periodicity of the burst cycle process, with the main features being the same as those shown in Fig.~\ref{fig:fiducial-Rt}.

As shown by the features at the leftmost edge of Fig.~\ref{fig:fiducial-Rt-extended}, the initial conditions are immediately unstable to the condition in Eq.~\eqref{eq:stability-criterion} and a burst cycle begins at $t=0$. This is simply a byproduct of the initial conditions in Eq.~\eqref{eq:init-density}, which do not reflect the pre-burst state calculated in Sect.~\ref{sec:pre-post-constraints} and shown in Fig.~\ref{fig:pre-post-examples}. For this reason, as explained in Sect.~\ref{sub:results-fiducial-model}, we discard this first outburst cycle and limit our analysis to $t\gtrsim3$\,kyr (as in Fig.~\eqref{fig:fiducial-Rt}).
We note that this is not the case with the models in Sect.~\ref{sub:vary-alpha}, as there we initialize the disk appropriately following the method in Sect.~\ref{sec:pre-post-constraints} and verified using our fiducial model in Appendix~\ref{apdx:pre-post-burst}.

\end{document}